\documentclass[aps,prr,reprint,superscriptaddress]{revtex4-1}

\usepackage[utf8]{inputenc}
\usepackage{bm}
\usepackage{graphicx}
\usepackage{siunitx}
\usepackage{hyperref}
\usepackage{amssymb}
\usepackage{amsmath}
\usepackage{xcolor}
\usepackage{comment}
\usepackage{tikz}
\newcommand{\GG}[1]{\ensuremath{\mathbf{#1}}}
\newcommand{\ind}[1]{_{\mbox{\scriptsize #1}}}
\newcommand{\ds}{\displaystyle}
\newcommand{\ex}			 {\ensuremath{\GG{e}_{x}}}			
\newcommand{\ey}			 {\ensuremath{\GG{e}_{y}}}
\newcommand{\ez}			 {\ensuremath{\GG{e}_{z}}}
\newcommand{\ombar}			 {\ensuremath{\overline{\omega}}}

\newcommand {\Derivee}[2]     {\ensuremath{\frac{\displaystyle \mathrm{d}#1}{\displaystyle \mathrm{d}#2}}}
\DeclareSIUnit\gauss{G}

\graphicspath{{figures/}}

\begin{document}
\title{
A versatile ring trap for quantum gases
}
\author{Mathieu de Go\"er de Herve}
\address{Laboratoire de physique des lasers, Universit\'e Sorbonne Paris Nord F-93430, Villetaneuse, France.}
\address{LPL CNRS UMR 7538, F-93430, Villetaneuse, France.}
\author{Yanliang Guo$^{1,2}$}
\author{Camilla De Rossi$^{1,2}$}
\author{Avinash Kumar$^{2,1}$}
\author{Thomas Badr$^{2,1}$}
\author{Romain Dubessy$^{1,2}$}
\author{Laurent Longchambon$^{1,2}$}
\author{H\'el\`ene Perrin$^{2,1}$}

\begin{abstract}
We report on the confinement of a Bose–Einstein condensate in an annular trap with widely tunable parameters. The trap relies on a combination of magnetic, optical and radio-frequency fields. The loading procedure is discussed. We present annular traps with radii adjusted between 20 and \SI{150}{\micro\metre}. We demonstrate the preparation of persistent flows both with a rotating laser stirrer and with a global quadrupole deformation of the ring. Our setup is well adapted for the study of superfluid dynamics.
\end{abstract}
\maketitle

\section{Introduction}
One of the main goals of the emerging field of atomtronics is to use the quantum properties of a well controlled atomic system to simulate another quantum system, or to measure with unprecedented accuracy some external parameter \cite{Amico2017}. In most of these situations, an atomic flow has to be generated and guided with minimum dissipation to preserve the coherence of the system \cite{Ryu2015}. In such a guide, a superfluid Bose–Einstein condensate (BEC) can simulate a charged current in a superconductor \cite{Caliga2017}, and with the presence of a repulsive barrier creates the equivalent of a Josephson junction \cite{Albiez2005,Kwon2020}. In that regard, the implementation of closed loops \cite{Gupta2005,Arnold2006,Sherlock2011,Pritchard2012,Navez2016,Marti2015,Bell2016,Heathcote2008,Ryu2007,Moulder2012,Ramanathan2011,Ryu2013,Wright2013a,Wright2013b,Eckel2014a,Ryu2020,Eckel2014b,Corman2014,Aidelsburger2017,Turpin2015,Eckel2018,Murray2013} is of paramount importance, because it allows to perform matter-wave interference, sensitive to rotation effects \cite{Gupta2005,Arnold2006,Sherlock2011,Pritchard2012,Marti2015,Navez2016,Bell2016,Helm2015}, to simulate with cold atoms the analogue of a SQUID \cite{Ryu2013,Ramanathan2011,Wright2013a,Wright2013b,Eckel2014a,Eckel2014b,Ryu2020} and to study out-of-equilibrium dynamics in systems with periodic boundary conditions \cite{Corman2014,Eckel2014b,Aidelsburger2017}. Moreover, the orbital angular momentum of the superfluid wavefunction is conserved for a circular loop, enabling the establishment of persistent currents in ring traps \cite{Ryu2007,Ramanathan2011,Moulder2012,Wright2013a,Wright2013b,Eckel2014a,Eckel2014b,Corman2014,Aidelsburger2017,Ryu2020} whose controlled coherent manipulation is promising to build elementary quantum logic blocks \cite{Solenov2010}.

One-dimensional (1D) rings are of particular interest because they can sustain solitons for interferometry \cite{Helm2015} and present different phase slip mechanisms relying on solitons \cite{Polo2019}. Reaching this regime requires that both trapping frequencies are larger than the thermal energy and the chemical potential, typically on the order of a few tens of nK in a quantum gas. Up to now, the experimental realizations of ring traps resulted in moderate trapping frequencies, the lowest of the two frequencies not exceeding \SI{300}{\hertz}, which makes the one-dimensional regime out of reach at typical temperatures. Increasing the trapping frequencies in ring traps is thus an important objective.

Radiofrequency-dressed adiabatic traps \cite{Zobay2001,Colombe2004a} have demonstrated kHz trapping frequencies with cylindrically symmetric geometries \cite{Merloti2013a}. In combination with a planar optical trapping potential, these potentials lead to a ring trap as proposed in \cite{Morizot2006} and first experimentally demonstrated in \cite{Heathcote2008} with thermal atoms, with potential access to large trapping frequencies. In this paper, we present the experimental realization of such an hybrid optical and magnetic ring trap, with the largest couple of radial and vertical trapping frequencies to the best of our knowledge. We demonstrate that this setup enables the preparation of persistent flows in a simple way \cite{Heathcote2008}, evidencing the superfluid character of the trapped gas.

This paper is organized as follows: after having described the principle of the trap and derived analytical formulas for its main parameters, we explain the loading procedure. Experimental results from oscillation frequencies and atoms losses are compared with theory and we discuss the feasibility of reaching the mean-field one-dimensional regime. Then we set the atomic ring into rotation by rotating either a magnetic trap deformation, or an optical repulsive stirrer. Time-of-flight density profiles of the rotating atoms demonstrate the superfluid character of the ring and we study the resulting matter-wave interference pattern as a function of the excitation frequency. By using an efficient reconnection technique to magnify the size of the central zero-density region for a rotating ring after a time-of-flight, we are able to detect up to a single quantum of circulation, and to observe the splitting of an unstable vortex of charge 3 into three singly charged stable vortices.

\section{Principle of the ring trap}
\label{sec:ring_principle}
\begin{figure}[t]
\begin{center}
\begin{minipage}{0.6\linewidth}
\centering
\includegraphics[width=0.95\linewidth]{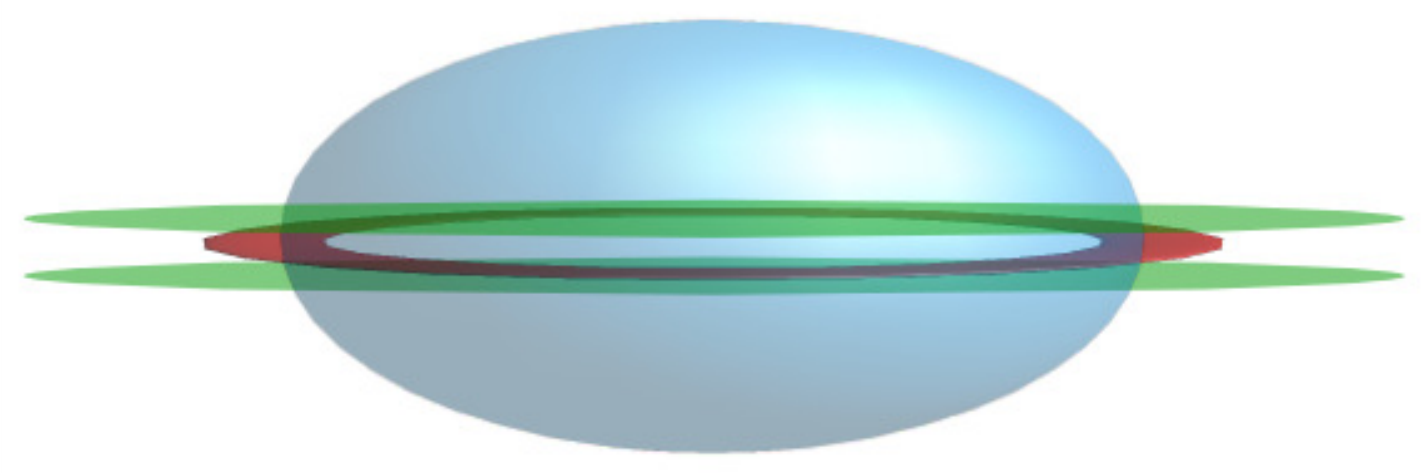}
\end{minipage}
\begin{minipage}{0.3\linewidth}
\centering
\includegraphics[width=0.95\linewidth]{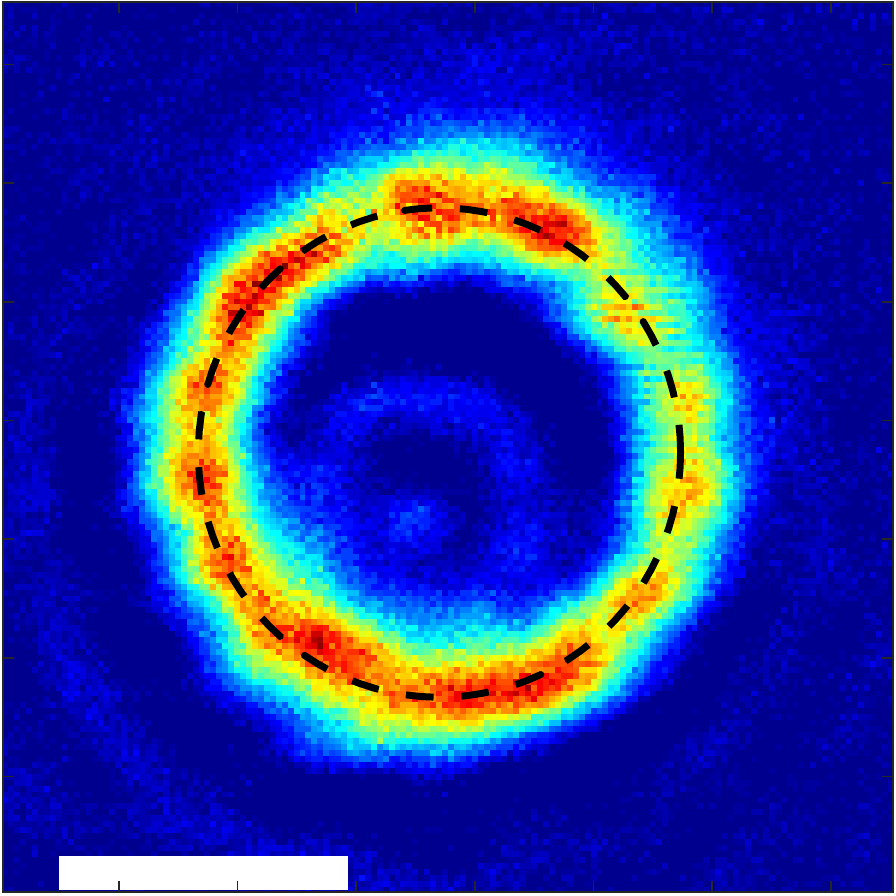}
\end{minipage}
\caption{\label{fig:ring_principle}Left: Principle of the ring trap. An ellipsoidal dressed quadrupole trap is cut at its equator by a double light-sheet. Right:  
Top view of an in situ density distribution of atoms confined in the ring trap.
The white bar has a length of \SI{50}{\micro\meter}.
}
\end{center}
\end{figure}

The ring potential, as described in \cite{Morizot2006,Heathcote2008}, relies on intersecting two independent trapping potentials: a planar optical trap, confining the atoms at a given height $z_0$, and a bubble-like trap produced by the dressing of these atoms with a radio-frequency (rf) field in the presence of a quadrupolar magnetic field \cite{Merloti2013a}, which ensures the radial confinement, see Fig.~\ref{fig:ring_principle}. As a result, the atoms are confined to the intersection of a horizontal plane and an ellipsoid with a rotational symmetry around a vertical axis, i.e. a circle. The confinement axes are vertical and radial when the plane intersects the ellipsoid near its equator, and trapping is ensured by the optical potential and the adiabatic potential, respectively.\footnote{Away from the equator, the trap transverse eigenaxes slightly deviate from vertical and radial, by an angle which in the results presented here is of a few degrees. In the following, we still identify the main trap axes as horizontal and vertical for simplicity.} The detailed measurements presented from Sec.~\ref{sec:meas_osc_freq} on were performed using ring traps with the optical trap confining the atoms at the equator of the ellipsoid, hence with vertical and horizontal main axes.

\subsection{Radial confinement}
\label{sec:radial}
The confinement of atoms to the surface of an ellipsoid has been described in detail elsewhere \cite{Merloti2013a,Perrin2017}. In short, it relies on dressing atoms of total angular momentum $F$ undergoing an inhomogeneous magnetic field $B(x,y,z)$ by a radio-frequency field of frequency $\omega$ \cite{Zobay2001,Colombe2004a,Garraway2016,Perrin2017}. If the adiabatic condition is fulfilled, the atoms follow an adiabatic potential with an avoided crossing at the positions in space where the Larmor frequency, $\omega_{0}(x,y,z)=|g_F| \mu_B B(x,y,z)/\hbar$, is resonant with the rf field frequency $\omega$. Here $g_F$ is the Land\'e factor and $\mu_B$ is the Bohr magneton. The locus of these resonant points $\omega_{0}(x,y,z)=\omega$ is an isomagnetic surface and corresponds to a potential minimum for low-field seeking adiabatic states. The adiabatic potential also depends on the local coupling between the rf field and the atom characterized by the Rabi frequency $\Omega\ind{rf}(x,y,z)$, which in turn depends on their relative orientation \cite{Perrin2017}. This introduces a modulation of the adiabatic potential inside the isomagnetic surface. Finally, the adiabatic potential in the dressed state mF reads within the rotating-wave approximation (RWA)

\begin{equation}
    \label{eq:pot_adia}
V\ind{ad}(\mathbf{r}) = m_F \hbar\sqrt{(\omega-\omega_0(\mathbf{r}))^2+\Omega\ind{rf}(\mathbf{r})^2}\,.
\end{equation}
In this work, we use rubidium 87 atoms in their $F=1$ ground state and we have $m_F=F=1$.

For a quadrupolar magnetic field with a vertical symmetry axis $z$, the local static magnetic field is of the form $b'(x \GG{e}_x + y \GG{e}_y - 2 z \GG{e}_z)$ ($b'>0$) and the resonant points lay on an ellipsoid defined by
\begin{equation}
x^2+y^2+4z^2 = r_0^2,
\end{equation}
where $r_0$ is related to the rf frequency $\omega$ through
\begin{equation}
r_0 = \omega/\alpha.
\label{eq:bubble_radius}
\end{equation}
Here, $\alpha = |g_F| \mu_B b'/\hbar$ represents the magnetic gradient $b'$ in the horizontal plane in frequency units. Like the magnetic field itself, the resonant ellipsoid is rotationally invariant around the $z$ axis. Choosing an rf polarization circular around the vertical $z$ axis preserves the axial symmetry, with a local value of the Rabi frequency which depends only on $z$: $\Omega\ind{rf}(z)=\Omega\ind{max}(1-2z/r_0)/2$  \cite{Merloti2013a,Perrin2017}. It is maximum at the bottom of the ellipsoid ($z=-r_0/2$) and half this value at the equator where $z=0$.

In the absence of other potentials, the trapping frequency transverse to the bubble $\omega_\perp$ is approximately given by \cite{Zobay2001,Perrin2017} $\omega_\perp=\alpha\ind{loc}(z)\sqrt{F\hbar/[M\Omega\ind{rf}(z)]}$ where $M$ is the atomic mass and $\alpha\ind{loc}(z)=\alpha\sqrt{1+12z^2/r_0^2}$ is the local magnetic gradient in a direction normal to the surface. At the equator we get a pure radial confinement with
\begin{equation}
\omega_r=\alpha\sqrt{2F\hbar/[M\Omega\ind{max}]}.
\label{eq:radial_freq}
\end{equation}

The center of the ellipsoid can be displaced by the addition of a homogeneous magnetic field. We use this property in the loading procedure from the dressed quadrupole trap to the ring trap.

In the experiment, the gradient $b'$ ranges from 59 to \SI{229}{\gauss\per\centi\metre}, leading to values of $\alpha/(2\pi)$ ranging from 4.1 to \SI{16}{\kilo\hertz\per\micro\metre}. Together with the rf frequency which we tune between \SI{300}{\kilo\hertz} and \SI{1.2}{\mega\hertz}, this determines the horizontal radius $r_0$ of the ellipsoid, see Eq.~\eqref{eq:bubble_radius}, which ranges between \SI{19}{\micro\metre} and \SI{290}{\micro\metre}. The rf source is produced by a homemade amplified Direct Digital Synthesiser (DDS) \cite{Merloti2013a}. The Rabi frequency can reach $\Omega\ind{max}=2\pi\times \SI{100}{\kilo\hertz}$ at the bottom of the bubble, which corresponds to $\Omega\ind{rf}=2\pi\times \SI{50}{\kilo\hertz}$ at the equator.

\begin{figure}[t]
\begin{center}
\begin{minipage}{0.29\linewidth}
\centering
\includegraphics[width=0.95\linewidth]{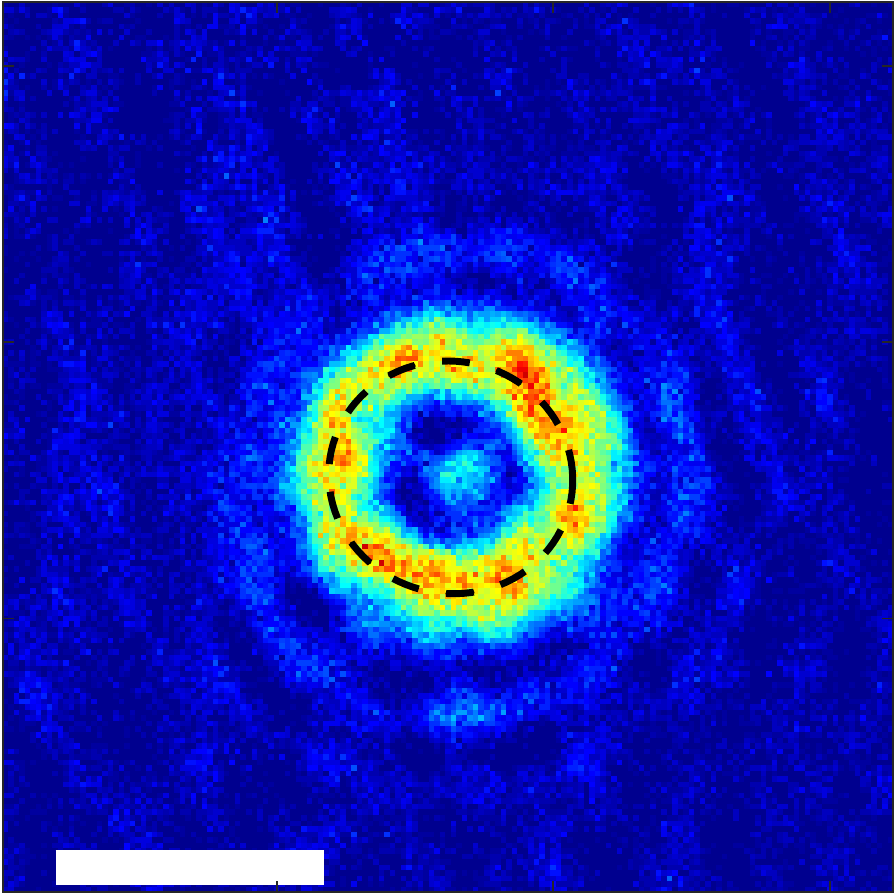} \end{minipage}
\begin{minipage}{0.29\linewidth}
\centering
\includegraphics[width=0.95\linewidth]{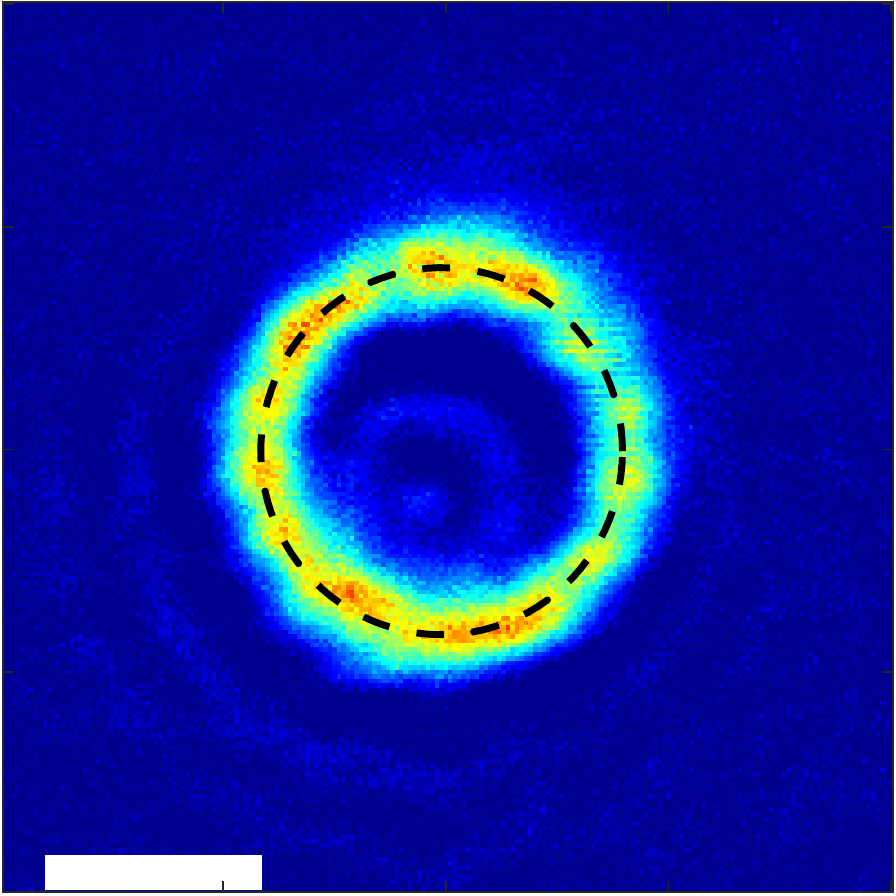} \end{minipage}
\begin{minipage}{0.29\linewidth}
\centering
\includegraphics[width=0.95\linewidth]{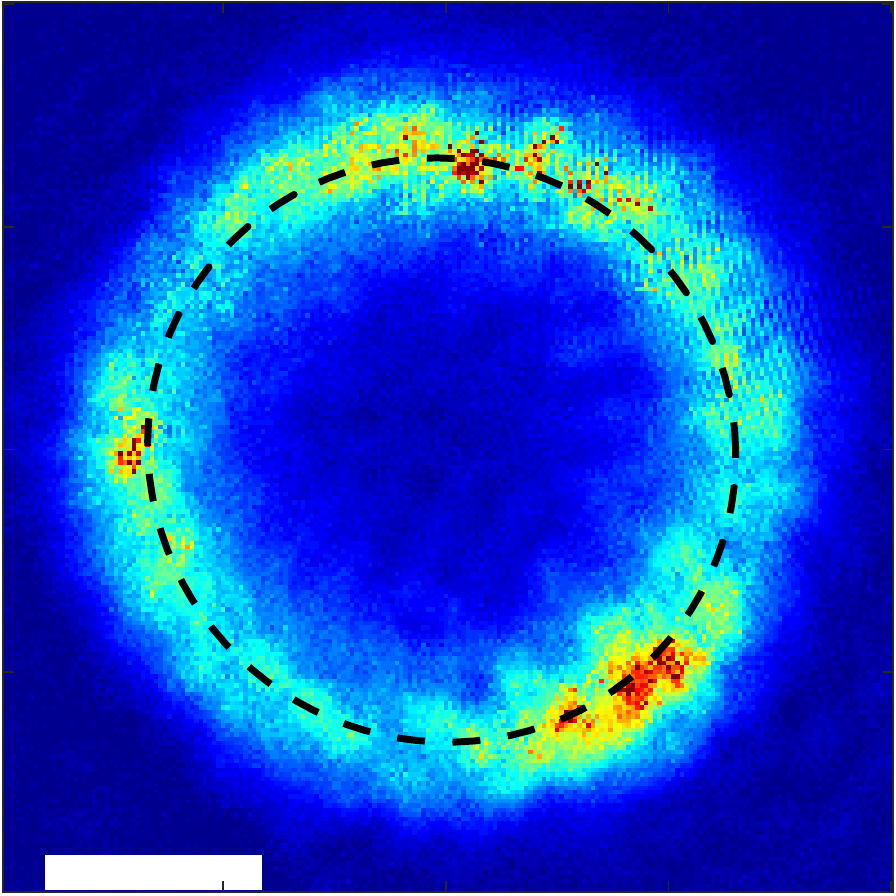} \end{minipage}
\,\\[2mm]
\begin{minipage}{0.29\linewidth}
\centering
\includegraphics[width=0.95\linewidth]{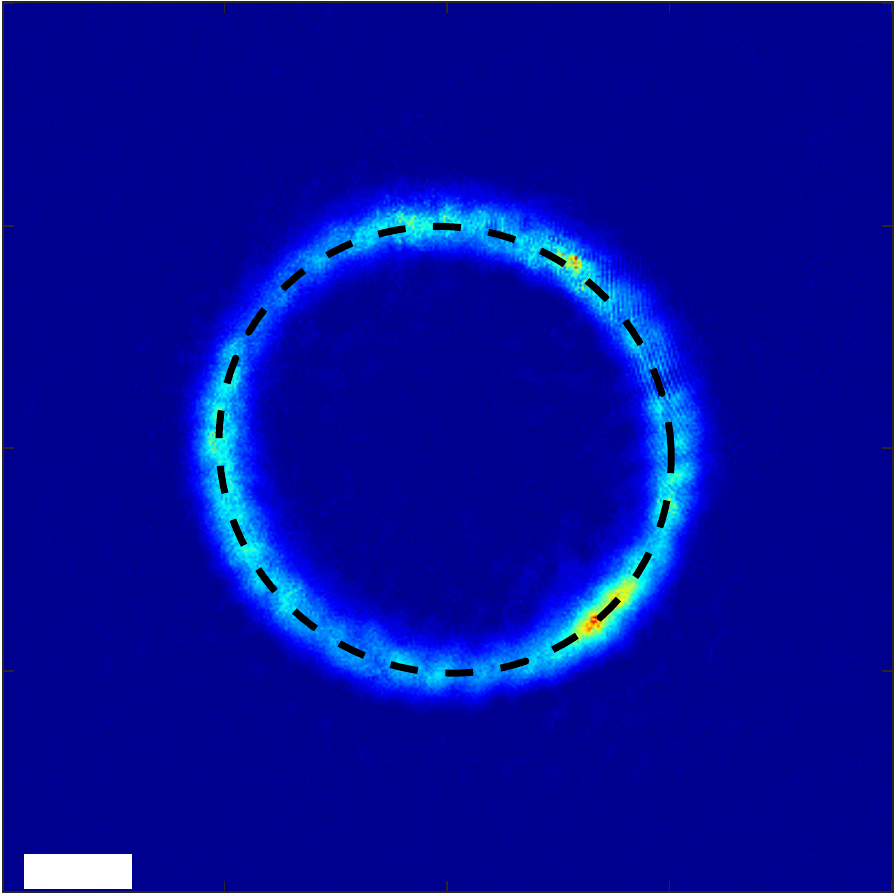} \end{minipage}
\begin{minipage}{0.29\linewidth}
\centering
\includegraphics[width=0.95\linewidth]{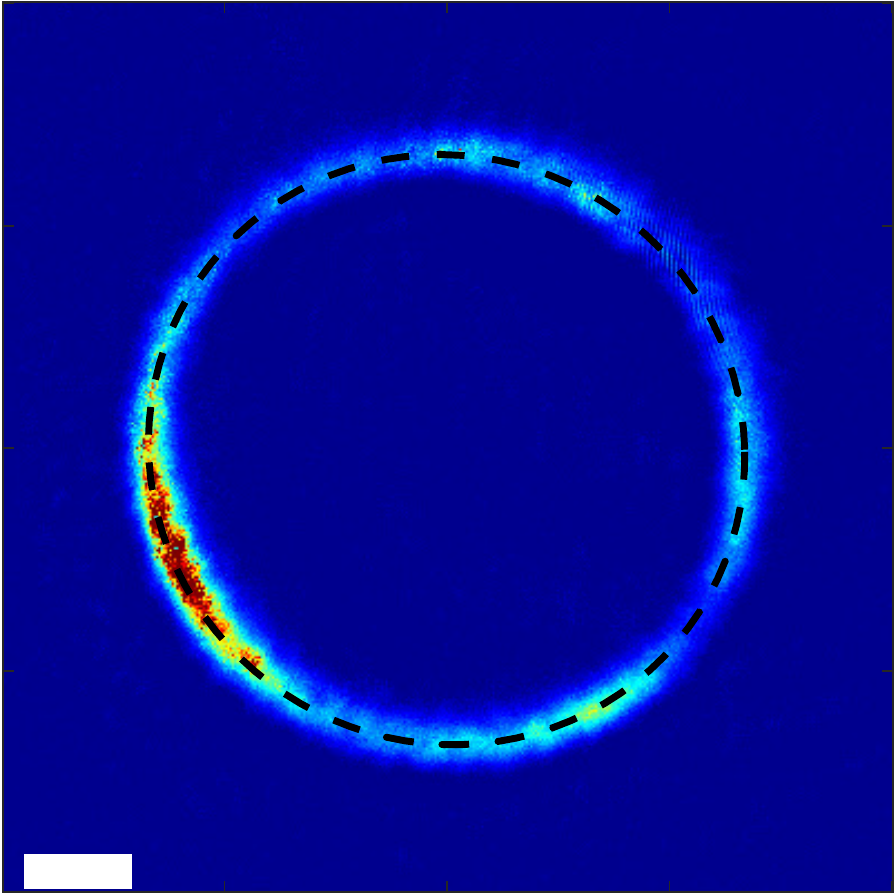} \end{minipage}
\begin{minipage}{0.29\linewidth}
\centering
\includegraphics[width=0.95\linewidth]{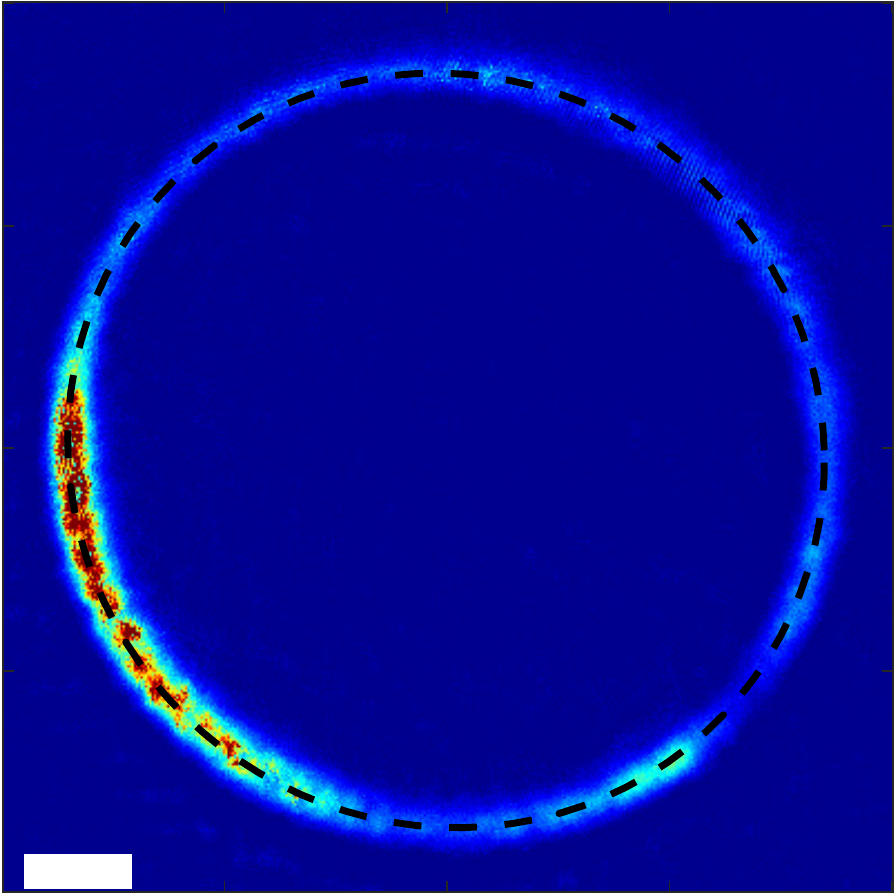} \end{minipage}
\,\\
\caption{\label{fig:single_ring}In situ pictures of ring-shaped gases obtained using various parameters. The dashed black line corresponds to the fitted radius. Top row: rings at the equator of the ellipsoid, with average $r_0=\SI{22.2}{\micro\metre}$, \SI{42.1}{\micro\metre} and \SI{67.7}{\micro\metre}. Bottom row: rings away from the equator obtained with the same ellipsoid with $r_0=\SI{244}{\micro\metre}$ and various heights $z_0$, giving $\rho_0=\SI{103.9}{\micro\metre}$, \SI{137}{\micro\metre} and \SI{174.5}{\micro\metre}. The anisotropy is at the level of $5\%$. The white bar in all images has a length of \SI{50}{\micro\meter}. In the top left picture, the central spot is a light diffraction effect.
}
\label{fig:ring_gallery}
\end{center}
\end{figure}

\subsection{Vertical confinement}
\label{sec:vertical}
The atoms are strongly confined in the vertical direction, at a given height $z_0$ of this ellipsoid, between two blue-detuned light sheets \cite{Heathcote2008,Meyrath2005,Smith2005}. This pair of light sheets results from the splitting of a laser beam of wavelength \SI{532}{\nano\metre} into two lobes by a holographic phase plate which produces a $\pi$ phase shift on the upper half of the beam. In the absence of the phase plate, the beam would be focused to a waist of $w_z=\SI{6}{\micro\metre}$ in the vertical direction and $w_x=\SI{200}{\micro\metre}$ in the horizontal direction. After the phase plate, the light field vanishes in an horizontal plane and the atoms can be confined near the zero field region at $z=z_0$. This produces an optical potential of the same form as in Ref. \cite{Smith2005} with a resulting maximum oscillation frequency in the harmonic approximation of \SI{2.72(2)}{\kilo\hertz} measured with a laser power of \SI{3}{\watt}, see Sec.~\ref{sec:meas_osc_freq}. The two waists are chosen to minimize the variation of this vertical oscillation frequency at fixed distance from the focal point as a function of the azimuthal angle, the optimal choice being a vertical Rayleigh length $\pi w_z^2/\lambda$ equal to $\sqrt{3}/2 w_x$.

In the experiment we typically use a reduced laser power of \SI{300}{\milli\watt}, leading to an estimated vertical oscillation frequency of \SI{850}{\hertz}. Even at this lower power, the confinement is very strong as compared to the effect of gravity and of the gradient of Rabi frequency $\Omega\ind{max}/r_0$ \cite{Morizot2007}, such that the position of the light minimum sets the potential minimum at $z=z_0$ to better than a micrometer.

The resulting ring has a radius approximately equal to
\begin{equation}
\rho_0 = \sqrt{r_0^2-4z_0^2},
\label{eq:ring_radius}
\end{equation}
which reaches its maximum value $r_0$ if the atoms are confined at the equator of the ellipsoid. Hence, the ring radius can be adjusted in three ways: shifting the height $z_0$, modifying the rf frequency $\omega$ or changing the magnetic gradient $b'$. The two last options also affect the underlying ellipsoid by changing the value of $r_0$.

In the case $z_0=0$ where the laser potential confines the atoms at the equator of the bubble, the ring trap has a radius $r_0$ and the vertical and horizontal confinement are independent, ensured by the optical potential and the adiabatic potential, respectively. In this case the radial frequency is approximately given by Eq.~\eqref{eq:radial_freq}. If $z_0\neq 0$, the ring radius is reduced following Eq.~\eqref{eq:ring_radius} and the trap eigenaxes slightly deviate from vertical and radial, with a reduced frequency in the direction closest to the horizontal and an increased frequency in the direction closest to the vertical.

Figure~\ref{fig:ring_gallery} shows examples of ring traps with various radii ranging from 22 to \SI{171}{\micro\metre}, illustrating the versatility of the hybrid ring trap.

\section{Loading procedure}
\label{sec:loading}
The optical alignment of the laser beam which creates the double light-sheet and the fine tuning of the experimental parameters are detailed in Appendix~\ref{sec:alignment}. In this section we describe the loading process from the dressed quadrupole trap into the ring trap.

The ring trap can be loaded from a connected ultracold atom cloud in several ways. Unlike Ref. \cite{Heathcote2008}, we start from a very anisotropic cloud already confined in the bubble-like dressed quadrupole trap \cite{Merloti2013a}. In this way, the gas is already very flat and a good mode matching with the strong vertical optical confinement is obtained.

Starting from the adiabatic potential at a given magnetic gradient, rf frequency and rf polarization, we first apply an additional homogeneous vertical magnetic field $B_z \GG{e}_z$, which shifts the trap upwards and allows us to align the bottom of the ellipsoid with the zero intensity plane of the double light sheet. A check of the alignment can be achieved by briefly shining the laser beam for \SI{100}{\micro\second} as the atoms are released from the trap in time-of-flight expansion: if the atoms are at the dipole trap minimum, the beam should not accelerate the cloud upwards nor downwards and its center-of-mass after time-of-flight should be at the same position than without the pulse. Once this alignment is performed, the atomic cloud is loaded into the dipole trap by ramping up the laser power within \SI{300}{\milli\second} to its maximum value, with a profile following a cosine branch in order to minimize excitations during this process.

\begin{figure}[b!]
\begin{center}
\includegraphics[width=\linewidth]{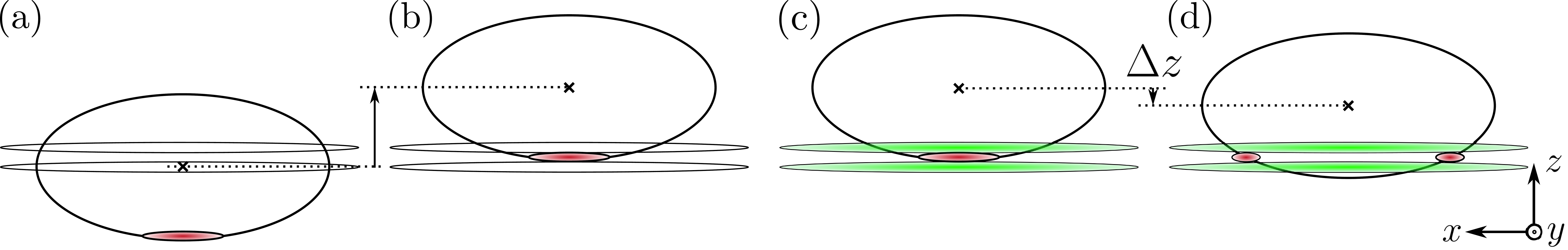}
\caption{Light sheet loading and ring formation procedure. In a first step (a-b), the bubble is lifted to align the atoms with the double light sheet, which is then turned on, with a propagation axis along $y$ (c). The bubble is then shifted downwards by $\Delta z=\Delta B_z/b'$, see Eq.~\eqref{eq:final_radius}, to generate a ring-shaped trap (d).
\label{fig:sheet_loading}}
\end{center}
\end{figure}

The height of the cloud is now fixed by the double light sheet. We obtain an annular gas by displacing the bubble downwards (Fig.~\ref{fig:sheet_loading}). To this aim, the value of $B_z$ is further modified by $\Delta B_z < 0$, which shifts the center of the isomagnetic ellipsoid downwards by $\Delta B_z/(2b')$. The atoms are pushed radially inside the adiabatic ellipsoidal potential and form a ring. The final ring radius reads
\begin{equation}
\rho_0 = \sqrt{r_0^2-(r_0 + \Delta B_z/b')^2} = r_0 \sqrt{2 \frac{|\Delta B_z|}{B\ind{res}} - \left(\frac{\Delta B_z}{B\ind{res}}\right)^2},
\label{eq:final_radius}
\end{equation}
where $B\ind{res}=b'r_0=\hbar\omega/(|g_F|\mu_B)$ is the resonant magnetic field.
The radius increases between zero and $r_0$ when $\Delta B_z$ is ramped between zero and $-B\ind{res}$.

\begin{figure}[t]
\begin{center}
\includegraphics[width=0.6\linewidth]{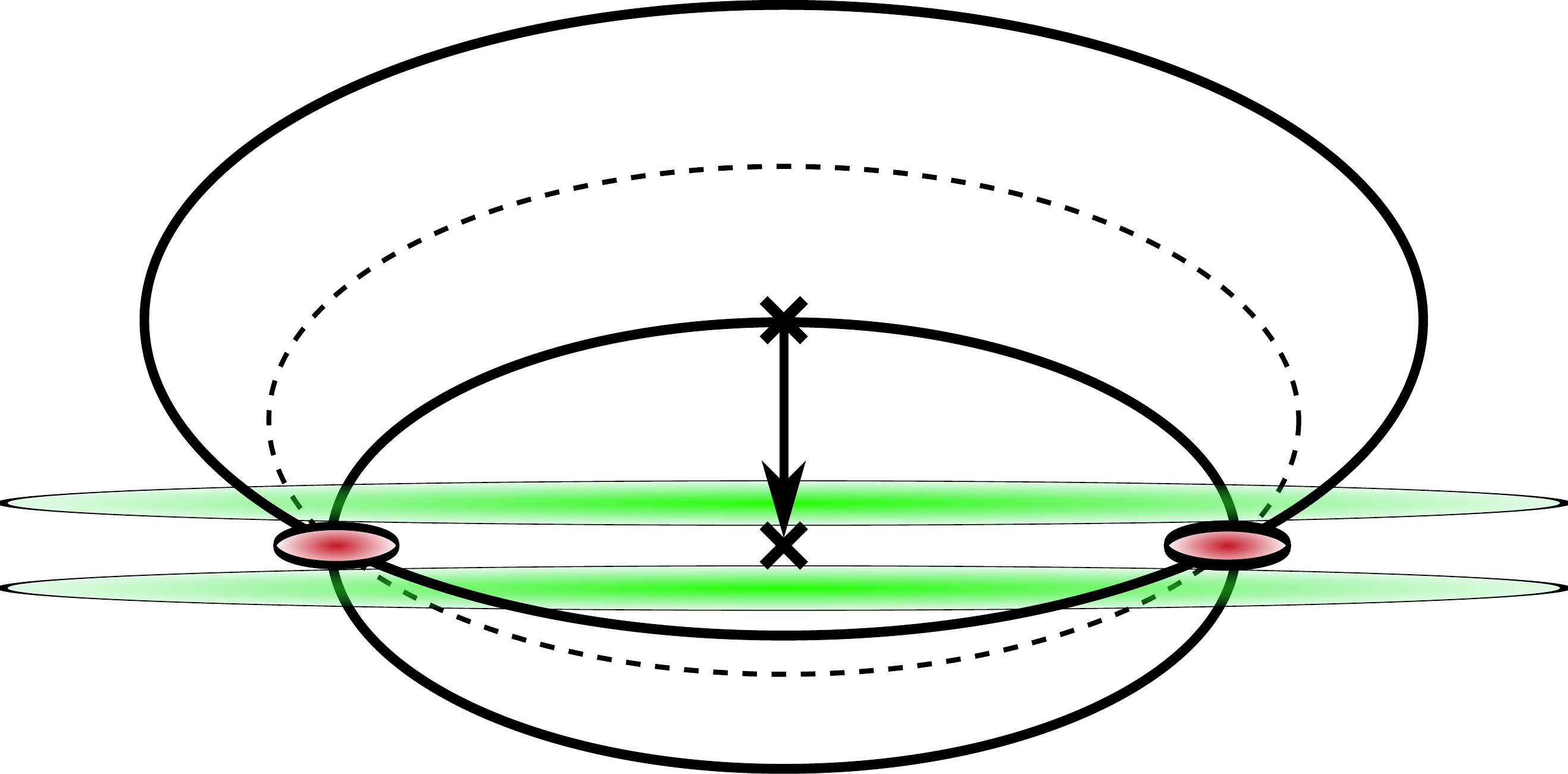}
\caption{Compression of the ring trap at constant radius: while the magnetic gradient increases, the size of the bubble decreases; the vertical bias is modified at the same time to keep the ring radius constant. At the end of this process, the ring is at the equator of the bubble with a larger radial frequency.}
\label{fig:ring_compression}
\end{center}
\end{figure}

In order to minimize heating in the ring trap due to magnetic field or beam pointing fluctuations, it is preferable to align the double light sheet with the equator of the bubble, where the ring radius is independent of the height to first order. Starting from the ring of Fig.~\ref{fig:sheet_loading}d, this is achieved at fixed ring radius by increasing the magnetic field while shifting the trap center, see Fig.~\ref{fig:ring_compression}. Using a larger magnetic gradient also leads to a smaller radius and a higher radial frequency, and thus a chemical potential larger than remaining potential defects due to light scattering.
At the end of this second step, the ring is located at the equator of the high-gradient ellipsoid.

\section{Characterization of the ring trap}
\label{sec:results}

We present here a full characterization of the ring trap in a typical configuration. We chose a radius $r_0$ significantly smaller than $w_x$ and the vertical Rayleigh length to minimize the azimuthal potential variations. The choice of a small radius also comes with a higher critical temperature and a larger chemical potential, which reduces the relative density fluctuations around the ring due to residual light scattering from the vacuum glass cell. Finally we align the optical trap with the equator of the ellipsoid to minimize the effect of beam pointing fluctuations.

The parameters of the ring described in this section are as follows: a dressing rf frequency of \SI{300}{kHz}, a Rabi frequency of $\Omega\ind{max}=2\pi\times \SI{100}{\kilo\hertz}$ at the bottom of the ellipsoid i.e. $\Omega\ind{rf}=2\pi\times \SI{50}{\kilo\hertz}$ at the equator. Data are presented with a current in the quadrupole coils set to either \SI{90}{\ampere} or \SI{100}{\ampere}, which gives $r_0=\SI{23}{\micro\metre}$ or $r_0=\SI{19}{\micro\metre}$, respectively. The laser power is between \SI{300}{\milli\watt} and \SI{3}{\watt}.

\subsection{Achieving condensation in the trap}
\label{sec:condensation}
\begin{figure}[t]
\begin{center}
\begin{tikzpicture}
\node at (-2,0.5) {\includegraphics[width=0.25\linewidth]{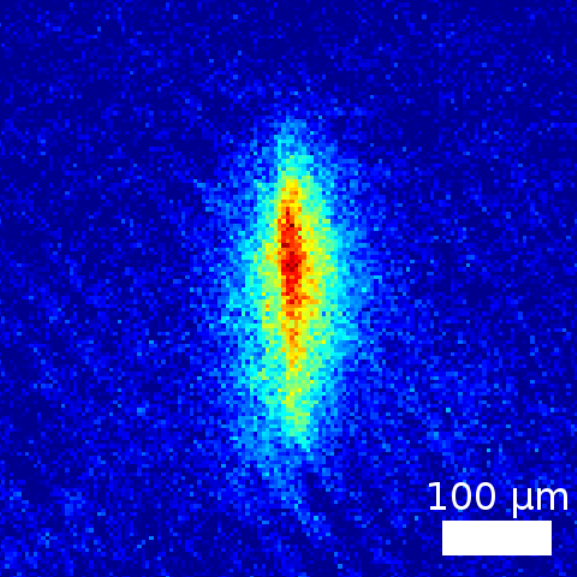}};
\node at (2.5,0) {\includegraphics[width=0.7\linewidth]{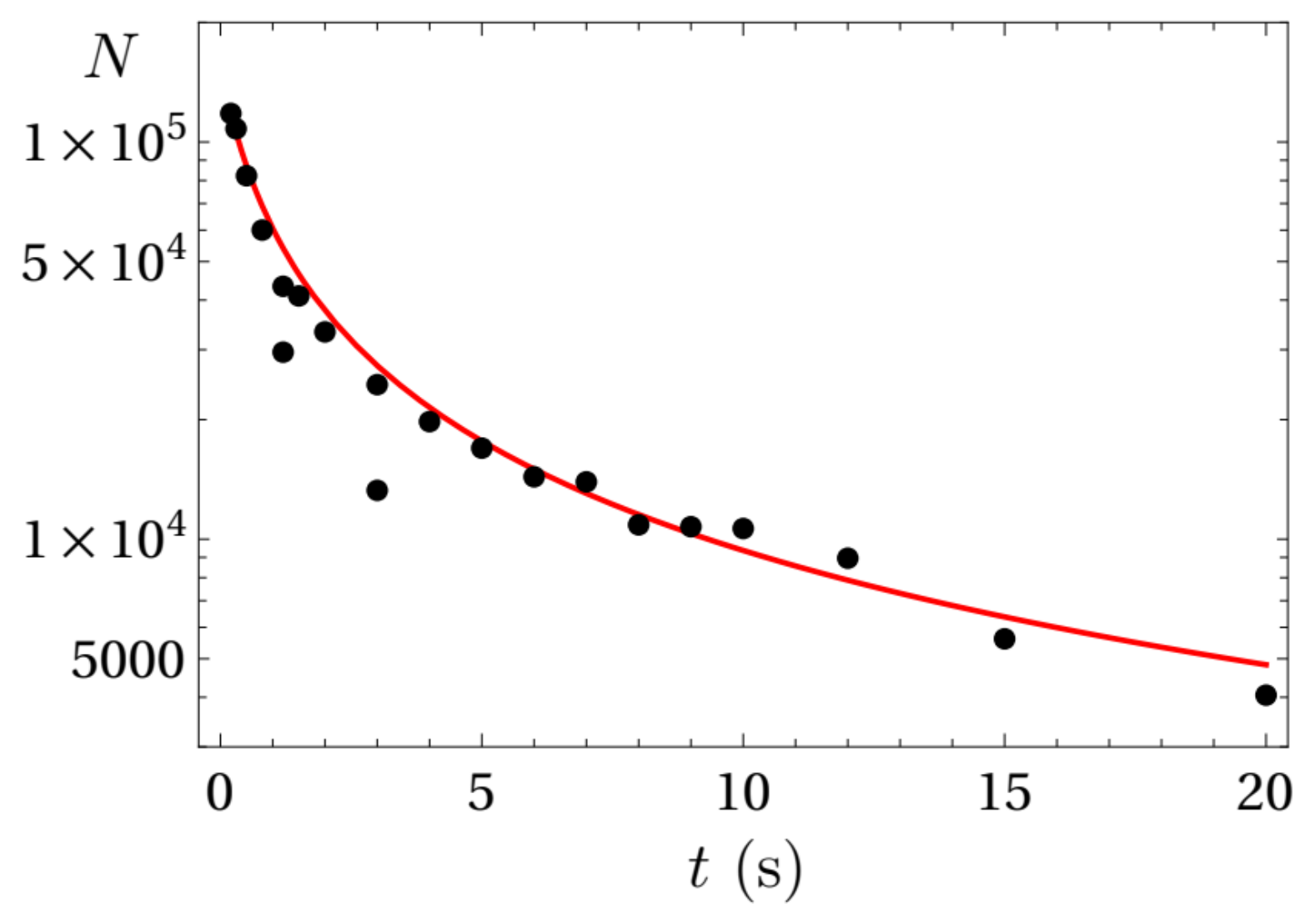}};
\node at (-3,2) {(a)};
\node at (-0.5,2) {(b)};
\end{tikzpicture}
\caption{a) Density distribution in the $xz$-plane of an annular condensate produced with $b'=\SI{186}{\gauss\per\centi\meter}$, $\omega=\SI{300}{\kilo\hertz}$ and $\Omega\ind{rf}=2\pi\times\SI{50}{\kilo\hertz}$ in the ring, as measured after a \SI{23}{\milli\second} time-of-flight expansion. The white bar has a length of \SI{100}{\micro\meter}. b) Atom number $N$ as a function of the hold time $t$ for an annular condensate with $b'=\SI{227}{\gauss\per\centi\meter}$, $\Omega\ind{rf}=2\pi\times\SI{56}{\kilo\hertz}$ measured in the ring and a laser power of \SI{600}{\milli\watt}. The red line is the prediction of Eq.~\eqref{eq:3b_Ring} using the first data point as $N(0)$, and no free parameter.}
\label{fig:Ring_lifetime}
\end{center}
\end{figure}

In order to control the temperature of the cloud, a radiofrequency knife \cite{Garrido2006} is applied during the whole experimental procedure. Its frequency is kept at least \SI{100}{\kilo\hertz} above the trap bottom during the initial formation of a dressed BEC, in order to keep a large number of atoms before loading the ring, such that the cloud is partly condensed with a large thermal fraction. During the whole ring loading procedure, the rf knife frequency is kept constant while the Rabi frequency decreases from \SI{100}{\kilo\hertz} to \SI{50}{\kilo\hertz} as the atoms move towards the equator, leading to an increased potential depth. Its role is then to ensure that the cloud temperature does not rise too much, while atom losses are kept low. At the end of the ring formation, the cloud has approximately $3\times10^5$ atoms, with a measured temperature of \SI{230}{\nano\kelvin}. The rf knife frequency is then lowered in a 2-step linear ramp, going from \SI{100}{\kilo\hertz} above the trap bottom to \SI{35}{\kilo\hertz} in \SI{100}{\milli\second} and then to \SI{25}{\kilo\hertz} in an additional \SI{200}{\milli\second}. At the end of the final rf knife ramp, we are left with approximately 1 to $1.5\times10^5$ atoms. The radial distribution as observed after a time-of-flight expansion (see Fig.~\ref{fig:Ring_lifetime}) is very peaked, as expected for the ground state in an annular gas where the momentum distribution is reminiscent from a Bessel $J_0$ function \cite{Berman2004,Ramanathan2011,Moulder2012,Murray2013}.

We find that the lifetime of the condensate in the ring trap is governed by three-body losses, as shown by the non exponential decay of Fig.~\ref{fig:Ring_lifetime}b. In our ring geometry, if we assume a Thomas-Fermi distribution in the radial and vertical directions, the atom number as a function of time is (see \ref{appendix:3D_losses}):
\begin{equation}
N(t)=\ds\frac{1}{\displaystyle K_{3}\left(\frac{M\omega_{r}\omega_{z}}{4\pi^{2} gr_{0}}\right)t+\frac{1}{N(0)}},
\label{eq:3b_Ring}
\end{equation}
where $g=4\pi\hbar^2a/m$ is the interaction constant and $a$ the scattering length, and $K_{3}=\SI{5.8+-1.9 e-42}{\meter\tothe{6}\per\second}$ \cite{Burt1997}. Using the expected trap parameters and the measured initial atom number as input, this formula predicts very well the observed time evolution of the number of atoms remaining in the trap as shown in Fig.~\ref{fig:Ring_lifetime}b.

\subsection{Oscillation frequencies in the trap}
\label{sec:meas_osc_freq}

\begin{figure}[b!]
\begin{center}
\begin{tikzpicture}
\node at (-2.2,0) {\includegraphics[width=0.45\linewidth]{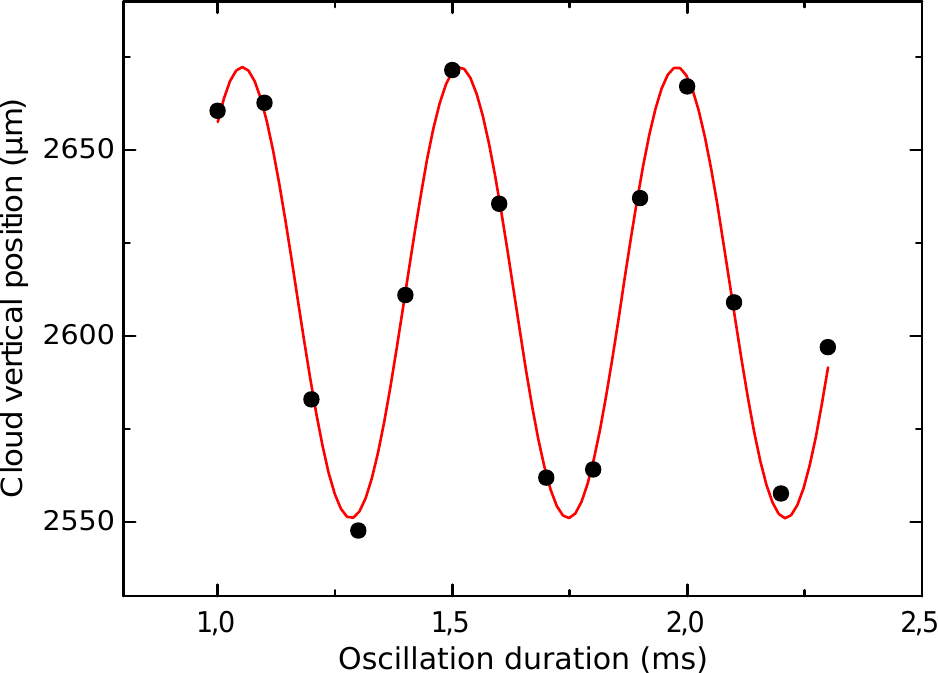}};
\node at (2.2,0) {\includegraphics[width=0.44\linewidth]{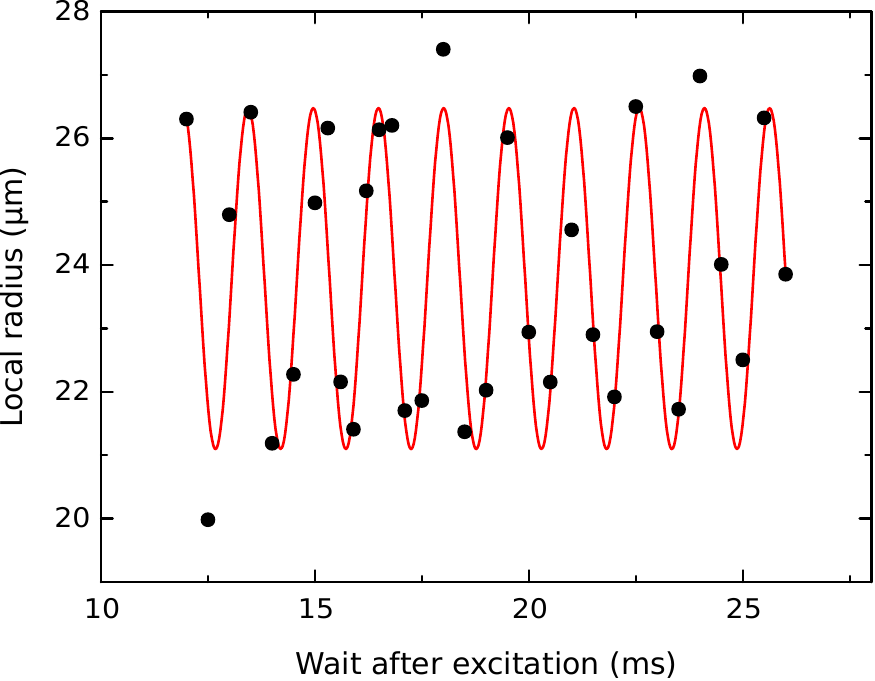}};
\node at (-3.9,1.6) {(a)};
\node at (0.2,1.6) {(b)};
\end{tikzpicture}

\caption{a) Closed circles: vertical position of the cloud after a sudden release from the double light sheet potential, for one third of the maximal laser power i.e. \SI{0.9}{\watt}, measured after \SI{23}{\milli\second} time-of-flight. A sinusoidal fit (red line) gives a frequency of \SI{1.46(1)}{\kilo\hertz}. b) Closed circles: average distance of a \SI{30}{\degree} wide section of the annular gas to the center-of-mass of the whole ring after excitation. The sinusoidal fit (red line) gives a frequency of \SI{656(3)}{\hertz}.\label{fig:frequencies_measurement}}
\end{center}
\end{figure}

As the ring is formed at the equator of the ellipsoid, the vertical oscillation frequency in the ring trap is set only by the double light sheet. We measure the trapping frequency in the center of the dipole potential alone by recording the oscillations of the center of mass. To this aim, we first load a dressed atomic cloud in between the two light sheets (steps (a) to (c) of Fig.~\ref{fig:sheet_loading}). The resonant ellipsoid is then shifted upwards to align the atomic spins with the bare magnetic state $m_F=-1$. This also drags the center of mass of the cloud slightly above the center of the dipole trap because of the magnetic force. The magnetic field is then turned off abruptly and the atoms oscillate freely in between the two light sheets for various holding times; finally we turn off the laser and measure the position of the cloud after a \SI{23}{\milli\second} time-of-flight. For a laser at full power we obtain a vertical trapping frequency of \SI{2.72(2)}{\kilo\hertz} (Fig.~\ref{fig:frequencies_measurement}a). In the ring trap, this maximal trapping frequency is slightly reduced because the atoms are not exactly at the center of the light sheet but at a distance $r_0$. For $r_0=\SI{23}{\micro\meter}$ this reduction is expected to be equal to $-2.6\%$, but for \SI{50}{\micro\meter} it reaches $-12\%$ and for \SI{100}{\micro\meter} $-40\%$.

We proceed in the same way to measure the radial oscillation in the trap. We first increase slowly the ring radius by sweeping the dressing frequency from 300 to \SI{308}{\kilo\hertz} in \SI{9.8}{\milli\second}, and then quickly bring back the radius to its initial value in \SI{0.2}{\milli\second}. This duration is long enough to ensure an adiabatic following of the dressed states but much shorter than the trapping period, thus preparing the atomic cloud out of equilibrium. The radius of the ring-shaped cloud starts to oscillate. After a given evolution time, the magnetic trap is switched off and the cloud expands in between the two light sheets for \SI{3}{\milli\second}. This short time-of-flight enhances the oscillation amplitude while preserving the annular shape of the cloud. We then take vertical absorption images and measure the cloud radius, see Fig.~\ref{fig:frequencies_measurement}b. We observe a small dephasing of the oscillation between the different angular sectors of the ring, due to a weak variation of the radial trapping frequency along the ring. We therefore average the value of the radius on twelve \SI{30}{\degree} angular sectors. For a ring with a gradient $\alpha=\SI{13}{\kilo\hertz\per\micro\metre}$ and a Rabi frequency \SI{46.9}{kHz} at the equator of the ellipsoid, the expected radial frequency is \SI{647.2}{\hertz}. 
We measure a mean radial oscillation frequency of $\omega_{r}=2\pi\times\SI{643(1)}{\hertz}$ with a 4\% peak-to-peak variation along the ring, ranging from 632 to \SI{656}{\hertz}, close to the calculated value.

\subsection{Towards quasi one-dimensional annular gases}
We now discuss the feasibility of reaching the one-dimensional regime in the ring trap, made possible by the large oscillation frequencies that can be reached in our system. The gas will enter this regime if its chemical potential and its temperature become comparable or lower than the two harmonic oscillator level splitting. The chemical potential of the ring-shaped gas can be estimated by its value in the three-dimensional Thomas-Fermi regime or in the quasi-condensate one-dimensional regime, given respectively by  \cite{Morizot2006}:
\begin{equation}
\mu\ind{3D}=\hbar\overline{\omega}\sqrt{\ds\frac{2 N a}{\pi r_{0}}},\qquad
\mu\ind{1D}=\hbar\overline{\omega}\ds\frac{ N a}{\pi r_{0}},
\label{eq:mu3D_ring}
\end{equation}
where $\overline{\omega} = \sqrt{\omega_r\omega_z}$ is the geometric average of the trapping frequencies. The most favorable configuration to have simultaneously $\mu<\hbar\omega_z$ and $\mu<\hbar\omega_r$ is to chose $\omega_r = \omega_z$.
From Eq.~\eqref{eq:mu3D_ring} we conclude that the gas will be in the quasi one-dimensional regime if
\begin{equation}
\ds\frac{2 N a}{\pi r_{0}}\lesssim 1,
\label{eq:1dcondition}
\end{equation}
which gives us a condition on the atomic linear density:
\begin{equation}
n_{1}=\ds\frac{N}{2\pi r_{0}}\lesssim\ds\frac{1}{4 a}.
\end{equation}
This density corresponds to 47 atoms per linear micrometer, which is small but still detectable with our apparatus. In the case of a ring with radius $r_0=\SI{23}{\micro\meter}$ as presented above, this corresponds to having less than 6800 atoms in the ring trap.

The temperature $T$ should also satisfy $k_BT<\hbar\omega_{r,z}$. The vertical oscillation frequency can be made larger than a kHz, such that it is straightforward to obtain $\omega_z = \omega_r$, the limiting factor being the radial oscillation frequency. The largest value for $\omega_r$ is reached for the maximum magnetic gradient $b'=228$~G$\cdot$cm$^{-1}$ or $\alpha/(2\pi)= \SI{16}{\kilo\hertz\per\micro\meter}$ and a small Rabi frequency. However the Rabi frequency cannot be chosen arbitrarily small to avoid Landau-Zener losses \cite{Burrows2017}. A good compromise is to set $\Omega\ind{rf}=2\pi\times\SI{40}{\kilo\hertz}$ at the equator, which corresponds to $\Omega\ind{max}=2\pi\times\SI{80}{\kilo\hertz}$ at the bottom of the ellipsoid. For this choice the radial trapping frequency reaches \SI{860}{\hertz}, and entering the quasi-1D regime requires a temperature below \SI{40}{\nano\kelvin}, which is not a stringent requirement. In particular, as the condition on the chemical potential implies that the atom number has to be less than $10^4$, a significant evaporative cooling in the ring trap from the initial atom number of $N=10^5$ will lead to both a temperature decrease and a lower atom number as requested. This evaporation generally leads to temperatures slightly below the transverse harmonic splitting \cite{Krueger2010}.

We also checked that in this regime, losses due to three-body recombination are not a problem. Once in the 1D regime, the atom number time evolution is governed by three-body losses and reads (see \ref{appendix:3D_losses}):
\begin{equation}
N(t)=\frac{1}{\sqrt{\displaystyle \frac{2}{3}K_{3} \left(\frac{M\overline{\omega}}{\pi r_{0} h}\right)^2 t+\frac{1}{N^2(0)}}}.
\label{eq:3b_Ring_1D}
\end{equation}
In our configuration, for a ring with $r_0=\SI{23}{\micro\meter}$ and $\overline{\omega}=2\pi\times\SI{860}{\hertz}$, with an initial atom number $N(0)=2\pi r_0/4a$, this process would lead to a half-life of about one minute. Moreover, in the one-dimensional regime the $K_{3}$ coefficient is expected to be reduced as the atoms enter the fermionized regime \cite{Laburthe2004}.

Entering the quasi-condensate one-dimensional regime seems within reach using the ring trap presented here. 
The crossover to the one-dimensional regime could be evidenced by the change in the time-dependant decay due to three-body losses (see Eqs.~\eqref{eq:3b_Ring} and \eqref{eq:3b_Ring_1D}), and the apparition of phase fluctuations along the ring \cite{Petrov2004}. 

\section{Preparation of persistent currents}
\label{sec:currents}

Ring traps can sustain persistent superfluid flows whose circulation along the ring is quantized \cite{Ryu2007,Moulder2012,Ramanathan2011,Ryu2013,Wright2013a,Wright2013b,Eckel2014a,Ryu2020,Eckel2014b,Corman2014,Aidelsburger2017,Murray2013}. The ring geometry is thus ideal to test the superfluid character of the trapped gas by producing atomic flows. 
In a frame rotating at a frequency $\Omega$, the ground state is expected to be a state with an integer phase winding $\ell$ and a circulation of the fluid velocity $\ell h/m$ where $\ell$ is the closest integer to $\Omega/\Omega_0$ and $\Omega_0=\hbar/(mr_0^2)$ is the elementary rotation corresponding to a circulation unity \cite{Wright2013a}.
Preparing a given circulation has been achieved by direct phase imprinting of the circulation \cite{Ryu2007,Ramanathan2011,Moulder2012,Kumar2018} or by rotating a potential anisotropy as for example a focused laser beam acting as a stirrer at a frequency $\Omega$ \cite{Wright2013a,Eckel2014a,Eckel2014b,Ryu2020}, or by a temperature quench \cite{Corman2014}. We have prepared superfluid flows following this second approach by rotating either a quadrupole deformation of the whole ring trap or a focused, blue-detuned, laser beam after the atoms have been loaded at low temperature in the ring trap.

The ring geometry used to prepare persistent flows is the following: the underlying adiabatic potential is realized with a magnetic gradient $b'=\SI{186}{\gauss\per\centi\meter}$ corresponding to $\alpha=\SI{13}{\kilo\hertz\per\micro\meter}$, an rf frequency of $\omega=2\pi\times \SI{300}{\kilo\hertz}$ and a Rabi frequency $\Omega\ind{max}=2\pi \times \SI{93.8}{\kilo\hertz}$ at the bottom of the bubble trap. We set the light sheet at the equator of the bubble as described in Sec.~\ref{sec:loading}. The local Rabi frequency in the ring trap is measured to be $\Omega\ind{rf}=2\pi\times \SI{48.0\pm 0.2}{\kilo\hertz}$, which corresponds to the field expected only \SI{0.3}{\micro\metre} below the equator. With these parameters, the ring radius is $r_0=\SI{23}{\micro\metre}$, which corresponds to an elementary rotation rate $\Omega_0 = 2\pi\times \SI{0.22}{\hertz}$. This gives us the typical scale for frequencies to apply to set the gas into rotation. In comparison, the sound velocity in the thin annular gas $c_{1D}=\sqrt{\mu/(2M)}$ \cite{Stringari1998}, computed from the chemical potential for $N=10^5$ atoms $\mu=h\times\SI{2.8}{kHz}$, is $\SI{2.5}{\milli\meter\per\second}$, which is reached in the ring for atoms rotating at a frequency  $\Omega=c_{1D}/r_0=2\pi\times\SI{18}{\hertz}$.

In order to observe the circulation of the rotating cloud after excitation, we perform a time-of-flight expansion by switching off the magnetic, rf and optical fields producing the ring trap. After a long enough free fall, a non-rotating ring exhibits a density maximum in the center, whereas in the presence of a nonzero circulation a central hole appears after expansion, which size is related to the initial circulation~\cite{Moulder2012,Murray2013}.

\subsection{Large rotations induced by a quadrupole deformation}
\begin{figure}[t]
\begin{center}
\begin{minipage}{0.4\linewidth}
\centering
\includegraphics[width=0.8\linewidth]{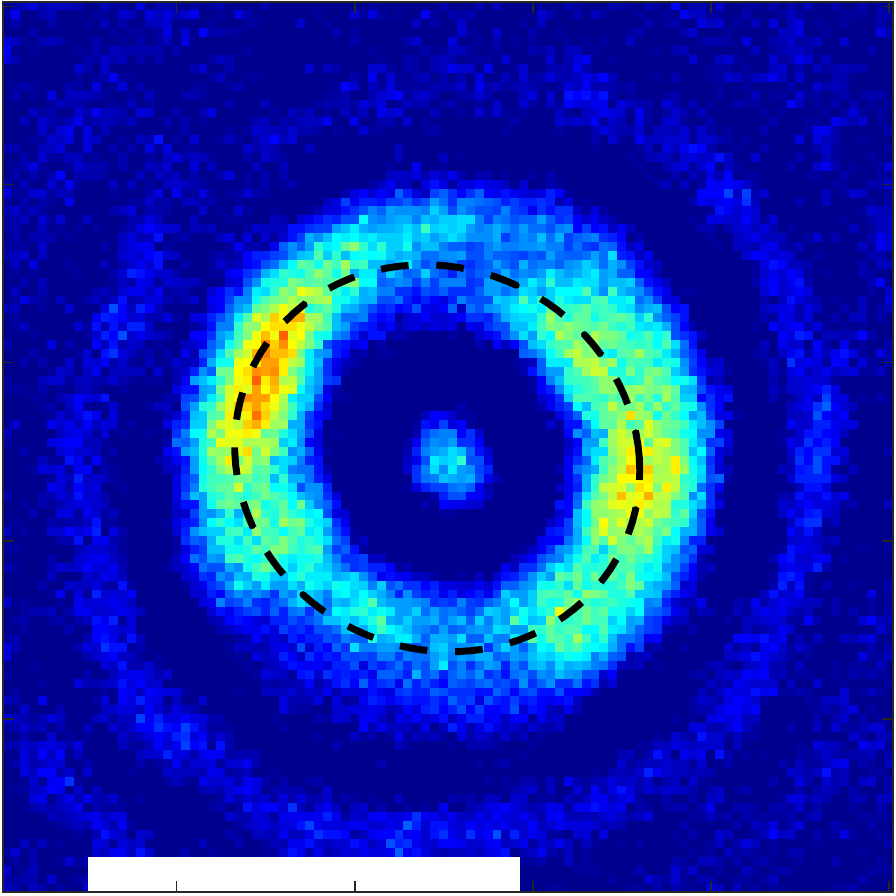}
\end{minipage}
\begin{minipage}{0.4\linewidth}
\centering
\includegraphics[width=0.8\linewidth]{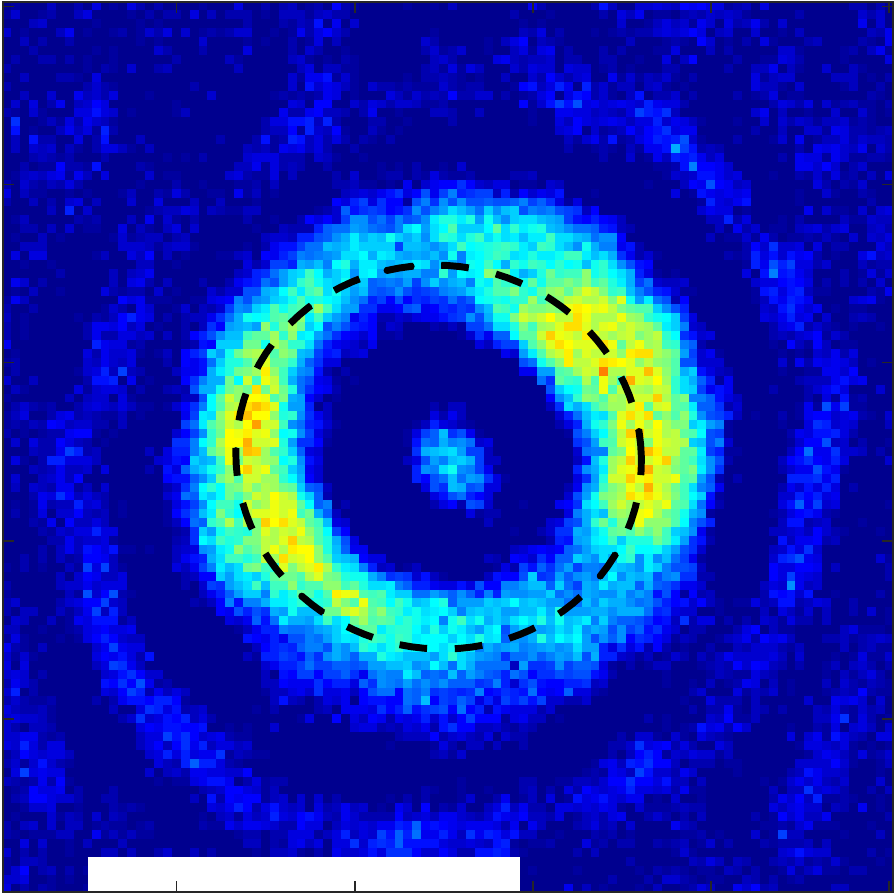}\\ 
\end{minipage}
\caption{\label{fig:deformation}
In situ absorption image of the atomic density with unbalanced rf polarization, for two choices of the main polarization axis. See  \ref{appendix:ringpotential} for detail.
The white bar in both images has a length of \SI{50}{\micro\meter}.
}
\end{center}
\end{figure}

As described in \ref{subsec:rf_polarization}, the rotational invariance around the ring can be broken using the rf antennas. We take advantage of this fine tuning of the ring geometry to induce a rotating quadrupole deformation of amplitude $\kappa$ around the annular gas by changing the phase and amplitude of the two antennas with horizontal axis. This technique has been used in the context of dressed traps with thermal atoms \cite{Heathcote2008, Sherlock2011} and recently allowed to reach the fast rotation regime in an adiabatic potential~\cite{Guo2020}. In this way, the in-plane potential on the resonant surface (i.e. for $r=r_0$) at a given azimuthal angle $\phi$ is made angle- and time-dependent:
\begin{equation}
V(r=r_0,z=0,\phi,t) = \hbar\Omega\ind{rf}\left[1+\kappa\cos{2(\phi-\Omega t)}\right]
\end{equation}
and two potential minima appear symmetrically around the ring circumference, rotating with angular frequency $\Omega$ (Fig.~\ref{fig:deformation}).

The experimental sequence is the following: $\SI{500}{\milli\second}$ after the ring formation, $2.5$ turns of a rotating deformation are applied with $\kappa=\SI{e-2}{}$, such that the modulation of the potential represents about $\pm 1/6$ of the chemical potential (see \ref{appendix:ringpotential}). The polarization is then tuned back to recover a rotationally invariant trap. After a free evolution in the ring trap, a time-of-flight expansion is performed followed by absorption imaging in the vertical direction.

\begin{figure}[h]
\centering
\begin{minipage}{0.15\linewidth}
\centering
\includegraphics[width=\linewidth]{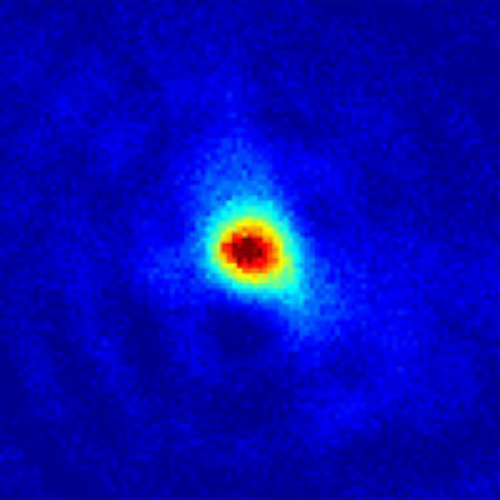}\\
(a)
\end{minipage}
\begin{minipage}{0.15\linewidth}
\centering
\includegraphics[width=\linewidth]{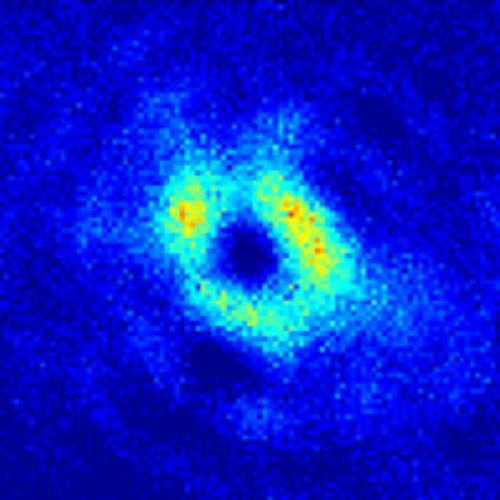}\\
(b)
\end{minipage}
\begin{minipage}{0.15\linewidth}
\centering
\includegraphics[width=\linewidth]{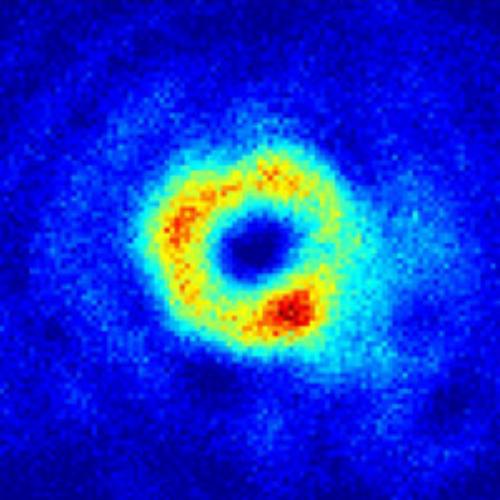}\\
(c)
\end{minipage}
\begin{minipage}{0.15\linewidth}
\centering
\includegraphics[width=\linewidth]{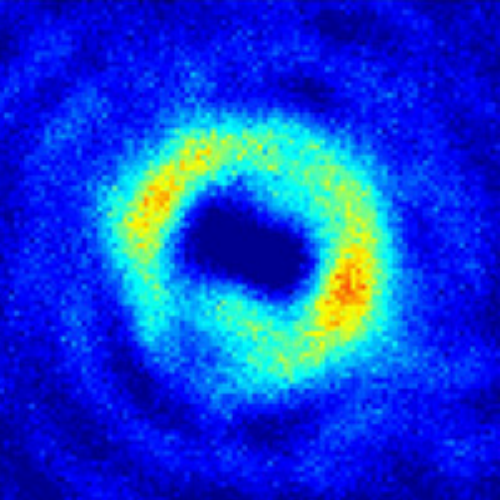}\\
(d)
\end{minipage}
\begin{minipage}{0.15\linewidth}
\centering
\includegraphics[width=\linewidth]{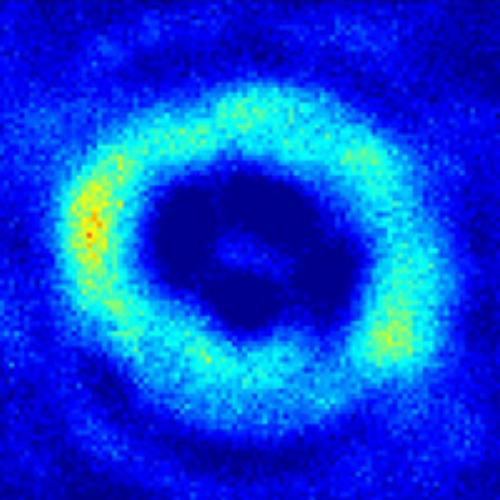}\\
(e)
\end{minipage}
\begin{minipage}{0.15\linewidth}
\centering
\includegraphics[width=\linewidth]{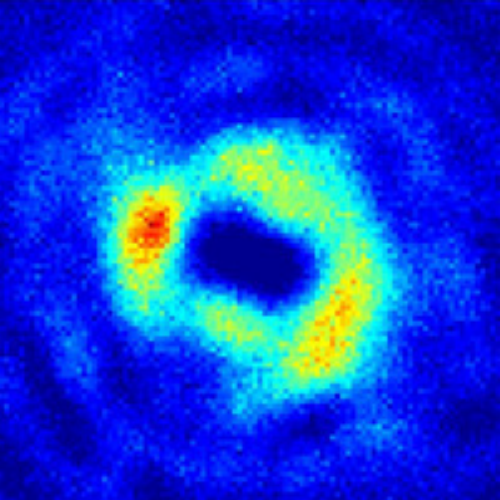}\\
(f)
\end{minipage}
\caption{Absorption pictures taken after a \SI{30}{\milli\second} time-of-flight expansion and various applied rotation rates of the quadrupole deformation. From (a) to (f), $\Omega/2\pi=\{0,6.8,13,15,18,30\}$~Hz. The image size is $100\times \SI{100}{\micro\metre^2}$. \label{fig:circulation_quad}}
\end{figure}

Figure~\ref{fig:circulation_quad} displays different experimental atomic density images obtained for various values of $\Omega$. For $\Omega=0$ the central density maximum is clearly observed, see Fig.~\ref{fig:circulation_quad}a. The presence of a central hole is a signature for a non-zero circulation in the ring. We have been able to observe this hole from $\Omega\simeq 2\pi\times \SI{6.8}{\hertz}$ to $\Omega\simeq 2\pi\times \SI{30}{\hertz}$, see Fig.~\ref{fig:circulation_quad}b-f. At larger rotation rates the excitation is less efficient and no circulation can be produced. The largest holes are produced with $\Omega\simeq 2\pi\times \SI{20}{\hertz}$, which suggests a resonant behaviour, as is the case for the rotating quadrupole excitation of connected gases \cite{Madison2001}. The linear velocity corresponding to this frequency is close to the speed of sound computed above. This method is well appropriate to produce fast rotating annular gases \cite{Guo2020}.

\subsection{Preparation of small quantized currents with a laser stirrer}
\label{sec:stirrer}
In a second series of experiments, we have prepared elementary excitations with another excitation method. A tailored optical potential can also break the rotational invariance in the trap and inject angular momentum into the ring. This technique has been used to rotate a connected cloud \cite{Madison2001} and also a ring trap \cite{Wright2013a} where it has successfully produced circulation quanta from $\ell=1$ to $\ell\gtrsim 12$ \cite{Murray2013}. In the latter case, it is predicted to rely on the density dip created by a rotating repulsive optical barrier which favours the entrance of vortices, eventually leading to an increase of circulation \cite{Piazza2009}.

In our case the stirring potential is a Gaussian beam of $\SI{7}{\micro\meter}$ waist and $\SI{532}{\nano\meter}$ wavelength, creating a local repulsive potential. Two acousto-optic modulators with orthogonal orientations enable dynamic control of the waist position along a circular path of radius $r\ind{stir}$ over an angle $\theta\ind{stir}$ during a time $t\ind{stir}$. Once the ring is formed, the stirrer power is ramped up to $\SI{1.4}{\milli\watt}$ in $\SI{500}{\milli\second}$ while rotating at fixed frequency $\Omega$, and keeps rotating during typically $t\ind{stir}=\SI{1}{\second}$ to $\SI{2}{\second}$ after which the stirrer power is ramped down to 0 again in $\SI{500}{\milli\second}$. As the typical Thomas-Fermi radius is $R=\SI{1.4}{\micro\meter}$, smaller than the waist, the situation differs from the case of a wide annulus described in \cite{Wright2013a} and we have found that in our case the most efficient configuration to impart circulation was to set $r\ind{stir}\simeq\SI{32}{\micro\meter}>r_0$. This corresponds to a light shift of \SI{0.8}{\kilo\hertz} at $r=r_0$, \textit{i.e} approximately $1/4$ of the chemical potential at this location.

In a first set of experiments where atoms are released directly from the ring trap for detection, the lowest stirrer frequency allowing to observe a central hole after time-of-flight is around five times the elementary rotation rate $\Omega_0$, meaning that either no rotation is imparted below this value, or our imaging resolution of $\SI{4}{\micro\metre}$ does not allow to observe smaller holes. In order to improve our detection threshold, we have adapted to our hybrid geometry a technique initially developed for a purely optical ring trap \cite{Ramanathan2011,Murray2013}.

Immediately after the stirrer is removed, the bubble trap is shifted up in the vertical direction in $\SI{100}{\milli\second}$ in order to reconnect the ring at the bottom of the bubble while maintaining the vertical optical confinement. This procedure reduces to its minimal value the radial trapping frequency $\omega_r$, preparing for a time-of-flight from a decompressed trap.  This decompression leads to an enhanced diameter of the hole size after time-of-flight in the case of a rotating ring. As soon as the ring is reconnected at the bottom, both magnetic and optical potentials are switched off and a vertical absorption imaging is performed after a $\SI{30}{\milli\second}$ time-of-flight. This technique allows us to observe a nonzero circulation for the lowest stirring frequencies around $0.7\Omega_0$, expected to create only one singly valued quantum of circulation.
We have also observed that for a larger rotation frequency of $\Omega=2\pi\times\SI{0.5}{\hertz}\approx 2.3\times\Omega_0$ and after a free evolution in the reconnected trap, the time-of-flight images show  a large single hole at evolution times shorter than $\SI{200}{\milli\second}$ and 2 to 3 smaller holes for longer times. This observation is compatible with initial persistent flows with $\ell=2$ and $\ell=3$, merged in the connected trap as a multiply charged vortex that finally decays into two and three $\ell=1$ single vortices as discussed and observed in \cite{Ryu2007,Moulder2012,Isoshima2007}.

\section{Conclusion}
In this paper, we describe the experimental realization of an hybrid magnetic and optical ring trap \cite{Morizot2006,Heathcote2008}.  Relying on an optical vertical confinement and a radial confinement from an rf-dressed bubble-shaped quadrupole trap at its equator, this well-controlled versatile ring trap allows a wide variety of radii and vertical and radial trapping frequencies, all independently tunable. In such a trap, we have achieved a ring-shaped Bose-Einstein condensate of radius $\SI{23}{\micro\meter}$ with no discernible thermal fraction, with trapping frequencies of $\SI{643}{\hertz}$ radially and a maximum of $\SI{2.72}{\kilo\hertz}$ vertically. We discuss the feasibility for this superfluid degenerate quantum gas to enter the long-sought one-dimensional regime with periodic boundaries and find that for current experimental values and atom number around 7000 the system should reach this quasi-condensate state. We also have set the atomic ring into rotation  with two different excitations, one purely magnetic relying on a quadrupole deformation of the bubble profile, and the other purely optical by means of a rotating focused beam. In the latter case especially, the observation of one multiply-charged vortex at short times after the excitation and two or three vortices at longer times is a strong signal that this trap sustains multivalued quantized circulation of the superfluid.
It would be interesting to investigate if this vortex splitting protocol enables to extract accurately the initial circulation in the ring. This hybrid ring is very promising for the study of 1D superfluid dynamics, for example the shock waves induced by rotation in the presence of a static barrier \cite{Dubessy2021}. The rotation can be imparted by a rotating defect as shown in the paper, but also by direct phase imprinting \cite{Kumar2018} onto the atoms at rest. Increasing again the ring confinement towards fermionization of the atoms could lead to NOON states more robust against decoherence \cite{Hallwood2010}, whereas  dressing the quadrupole static magnetic field with multiple rf frequencies \cite{Harte2018} allows the implementation of multiple concentric rings. Tunneling between these rings is expected to build a macroscopic quantum superposition of several rotating BECs \cite{Brand2009}. Moreover, the recombination technique, starting from a controlled large phase winding in the ring, could ease the way to the production of quantum Hall states as proposed in \cite{Roncaglia2011}.

\begin{acknowledgments}
We thank Thomas Bourdel for providing us with the phase-plate necessary to create the double light sheet beam.
LPL is UMR 7538 of CNRS and Sorbonne Paris Nord University. We acknowledge financial support from the ANR project SuperRing (Grant No. ANR-15-CE30-0012) and from the R\'egion \^Ile-de-France in the framework of DIM SIRTEQ (Science et Ing\'enierie en R\'egion \^Ile-de-France pour les Technologies Quantiques), project DyABoG.
\end{acknowledgments}

\appendix
\section{Potential along the ring in the case of an arbitrary polarization}
\label{appendix:ringpotential}

We consider a dressing field with arbitrary complex polarization $\boldsymbol{\epsilon}=\cos\Theta \cos\Theta_{z}\ex+ 
\sin\Theta\cos\Theta_{z}e^{i\Phi}\ey+\ 
\sin\Theta_{z}e^{i\Phi_{z}}\ez$. We want to compute the local rf coupling in the ring trap, that is, at the equator of the bubble. Defining the position with its azimuthal angle $\phi$, the quadrupole field orientation writes: $\GG{u}=\cos\phi\ \ex+\sin\phi\ \ey$. 

Taking into account the sign $g_F<0$ of the Land\'e factor in the $F=1$ ground state of rubidium 87, the local Rabi frequency can be written as \cite{Perrin2017}:
\begin{widetext}
\begin{eqnarray}
|\Omega\ind{rf}(\GG{r})| &=& \ds\frac{\Omega\ind{max}}{2}\sqrt{1-|\boldsymbol{\epsilon}\cdot\GG{u}|^{2}+|\boldsymbol{\epsilon}\times\GG{u}|^{2} - 2 i \GG{u}\cdot(\boldsymbol{\epsilon}\times\boldsymbol{\epsilon}^{*})}\\
&=&\ds\frac{\Omega\ind{max}}{2}\sqrt{1+\sin^2\Theta_z-A_\pi(\phi)\cos^2\Theta_z +2A_{2\pi}(\phi)\sin(2\Theta_z)}\nonumber
\end{eqnarray}
\end{widetext}
where we have defined
\begin{equation}
    A_\pi(\phi) = \cos(2\Theta)\cos(2\phi)+\cos\Phi\sin(2\Theta)\sin(2\phi)
\end{equation}
and
\begin{equation}
    A_{2\pi}(\phi) = \sin\Theta\sin(\Phi-\Phi_z)\cos\phi+\cos\Theta\sin\Phi_z\sin\phi.
\end{equation}
The choice of a $\sigma_-$ polarization along $z$, i.e. $\Theta=\pi/4$, $\Theta_z=0$ and $\Phi=-\pi/2$, gives $A_\pi=0$ and $\sin(2\Theta_z)=0$ and thus a uniform rf coupling with Rabi frequency $\Omega\ind{rf}=\Omega\ind{max}/2$. The expressions of $A_\pi$ and $A_{2\pi}$, periodic with a period $\pi$ and $2\pi$ as a function of $\phi$, respectively, help to understand the effect of a small polarization misalignment with respect to the ideal case:

(i) If the polarization lies in the horizontal plane ($\Theta_z=0$) but is elliptical ($\Theta\neq 0$ and/or $\Phi\neq -\pi/2$), $A_{2\pi}=0$ but $A_\pi\neq 0$ and the local Rabi frequency is modulated with a period $\pi$. The density in the ring presents a ``double-moon'' asymmetry.

(ii) If the polarization has a non-zero component along the axis of the quadrupole ($\Theta_{z}\neq0$), $A_{2\pi}$ is responsible for a modulation with period $2\pi$, which leads to a lateral imbalance in the ring with an amplitude linear in $\Theta_{z}$.

In the particular case of a polarization that is circular but slightly tilted compared to the quadrupole axis due to a tilt of the antennas ($\Theta=\pi/4$, $\Phi=-\pi/2$, $\Theta_{z}\ll1$), $A_\pi=0$ and $A_{2\pi}=-\cos(\phi+\Phi_z)/\sqrt{2}$. To first order in $\Theta_z$, the local rf coupling simplifies to:
\begin{eqnarray}
\Omega\ind{rf}(\phi)&\simeq&\ds\frac{\Omega\ind{max}}{2}\sqrt{1-\ds\frac{4}{\sqrt{2}}\Theta_{z}\cos(\phi+\Phi_{z})}\\
&\simeq&\ds\frac{\Omega\ind{max}}{2}\left[1-\sqrt{2}\Theta_{z}\cos(\phi+\Phi_{z})\right],
\label{eqCouplingTiltAntenna}
\end{eqnarray}
and we find a modulation of the Rabi frequency with a peak to peak amplitude $\sqrt{2}\Omega\ind{max}\Theta_{z}$ whose orientation is determined by the phase of the $z$ component of the polarization. Since we work with $\Omega\ind{max}=2\pi\times\SI{100}{\kilo\hertz}$, even a very small angle $\Theta_{z}$ can lead to a modulation whose amplitude will be on the order of a few kilohertz, see \ref{subsec:rf_polarization}.

\section{Fine tuning of the potential}
\label{sec:alignment}

\subsection{Laser beam}
\label{subsec:laser_beam}
\begin{figure}[t]
    \centering
\includegraphics[width=\linewidth]{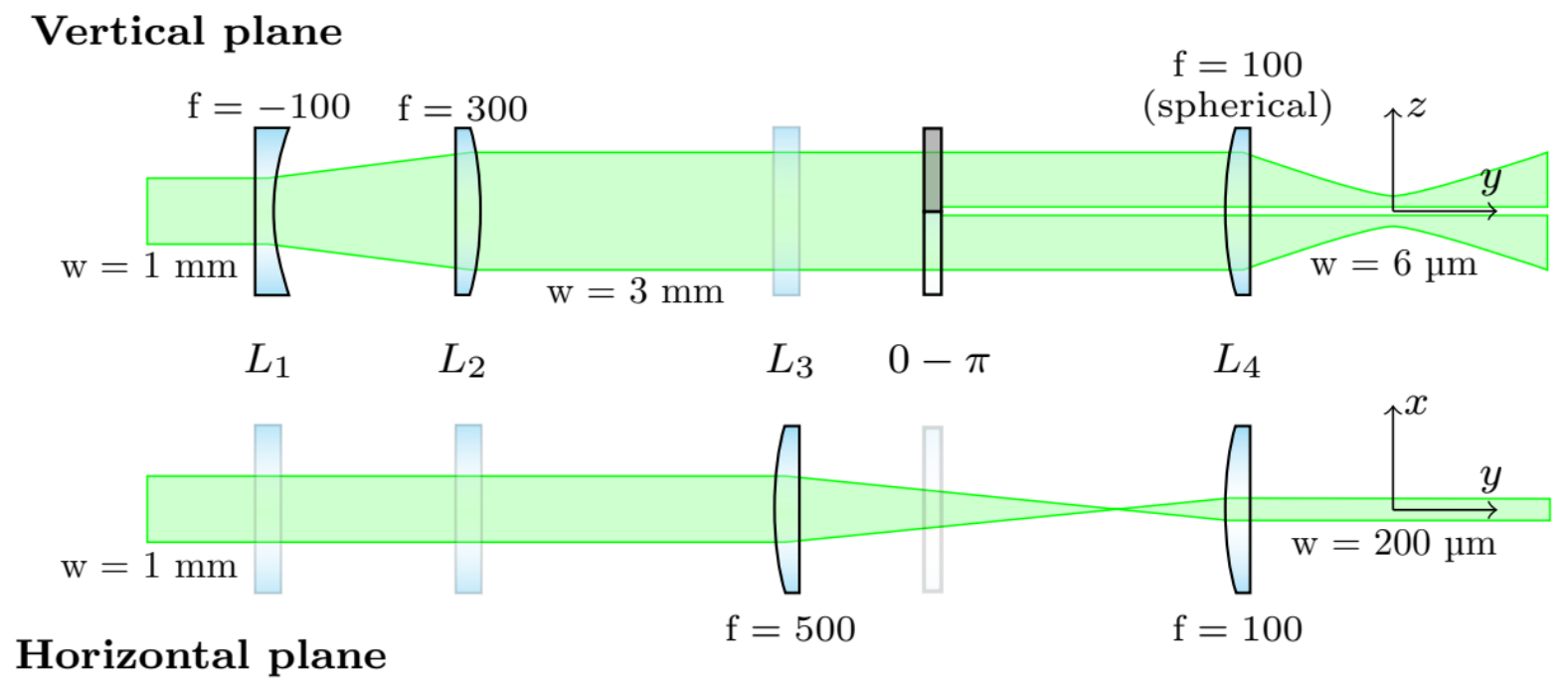}\\
    \caption{Optical setup used to shape the double light sheet beam. Top: beam profile in the vertical $yz$ plane. Bottom: beam profile in the horizontal $xy$ plane.
    \label{fig:sheet_principle}}
\end{figure}

The double light sheet responsible for the vertical confinement along $z$ requires a careful alignment in order to ensure that the resulting potential is minimum in an horizontal plane. The optical system consists in four lenses $L_1$ to $L_4$, the first three being assembled in a common cage system, and a $0-\pi$ phase plate placed between $L_3$ and $L_4$ (see Fig.~\ref{fig:sheet_principle}). The beam propagates along the $y$ axis. The two first lenses are cylindrical and are used to expand the initial isotropic Gaussian beam of waist ($1/e^2$-radius) 1~mm in the vertical direction by a factor 3, such that $w_z=3$~mm after these two lenses. The third lens $L_3$ is cylindrical, oriented along the horizontal axis, and together with the fourth spherical lens $L_4$ it provides a telescope reducing the horizontal beam waist by a factor 5, such that the beam is collimated in the horizontal direction to a waist $w_x=\SI{200}{\micro\metre}$. $L_4$ (focal length 100~mm) also focuses this beam in the vertical direction to a final waist of $w_z=\SI{6}{\micro\metre}$ at the position of the atoms in the absence of the $0-\pi$-phase plate. The position and orientation of these lenses and phase plate are critical. The cylindrical lenses $L_1$ to $L_3$ are placed in micrometric rotating stages, $L_4$ is mounted on a micrometric xyz translation stage, and the phase plate vertical position is also adjusted with a micrometric translation stage.

Once the atoms have been loaded between the two light sheets (see Fig.~\ref{fig:sheet_loading}c of Sec.~\ref{sec:loading}), the horizontality of the trapping plane (pitch and roll angles) is optimized. Firstly, in order to cancel the pitch angle, the dressed quadrupole trap is switched off, leaving the atoms expand inside the optical potential alone, and an absorption picture is taken along the vertical axis. Any non-zero pitch angle leads to a drift of the cloud center-of-mass along $y$ inside the optical potential due to gravity. This drift is suppressed by optimizing the vertical position of the lens $L_4$. Secondly, the effect of the roll angle is measured with an absorption picture taken along the horizontal $y$ axis, after a 23~ms free fall expansion of the cloud as both the optical potential and the bubble trap are switched off. The cloud after expansion is very anisotropic in the direction of the strong optical confinement. The roll angle is optimized to ensure that the main axis of the expanding cloud is indeed vertical. This is done by rotating simultaneously the three cylindrical lenses $L_1$-$L_3$ in their micrometric rotation mounts. Finally, the horizontal position of the laser beam is adjusted by displacing the lens $L_4$ along $x$ in order to maximize the cloud vertical size after this time-of-flight expansion. This ensures that the vertical trapping frequency is maximum at the location of the atoms, which is obtained at the center of the Gaussian horizontal profile where the intensity is maximum.

\subsection{Phase plate}
\label{subsec:phase_plate}
The vertical position of the phase plate is also very critical: if the light sheet beam is not split into two strictly equal parts the light field will not vanish at the center of the beam, leading to a central barrier in the horizontal plane. It can be adjusted with a precision of \SI{20}{\micro\meter} by looking from above at the expansion of the gas inside the optical potential alone. If the phase plate is not well-centered, the cloud is split into two parts that are pushed away from the center.

\subsection{Rf polarization}
\label{subsec:rf_polarization}

Finally, it is crucially important that the polarization of the rf field  belongs to the horizontal plane. Indeed, even a very small misalignment angle can lead to a large potential inhomogeneity on the scale of the chemical potential. While at the bottom of the ellipsoid the Rabi frequency has a small effect on the oscillation frequencies in the harmonic approximation \cite{Merloti2013a,Perrin2017}, once the atoms are at the equator of the ellipsoid gravity plays no role anymore and the potential landscape itself is given by the variations of the Rabi frequency with the azimuthal angle $\phi$: $V(r_{0},\phi)=\hbar\Omega\ind{rf}(\phi)$.
Any inhomogeneity in $\Omega\ind{rf}$ of a few percents will have a large impact For example, for a perfectly circularly polarized rf field, a tilt $\Theta_{z}\ll 1$ of the polarization plane leads to a potential difference of $\hbar\times \sqrt{2}\Omega\ind{max}\Theta_{z}$ between the two extreme positions in the ring (see \ref{appendix:ringpotential}). With our typical Rabi frequency $\Omega\ind{max}/2\pi=\SI{100}{\kilo\hertz}$, a \SI{1}{\degree} mismatch already leads to an inhomogeneity in the potential of almost $\SI{2.5}{\kilo\hertz}$, comparable to $\mu/h$.

Three antennas are thus required to produce the dressing field with a fully controlled polarization, even if the antennas axes are not strictly orthogonal. In practice, two main dressing antennas are used to create an rf field in a plane close to horizontal \cite{Merloti2013a}. A third antenna with vertical axis allows us to fine tune the polarization plane\footnote{The two main antennas are square-shaped with a \SI{16}{\milli\metre} side, made of ten loops of \SI{ 0.71}{\milli\metre} diameter copper wire. They are located at about \SI{10}{\milli\metre} from the atoms. The third antenna with vertical axis has a $11 \times \SI{6.5}{\centi\meter}$ rectangular shape, is made of four loops of copper wire and placed approximately \SI{5}{\centi\metre} below the atoms.}.

\begin{figure}[h]
\begin{center}
\begin{minipage}{0.19\linewidth}
\centering
\includegraphics[width=0.95\linewidth]{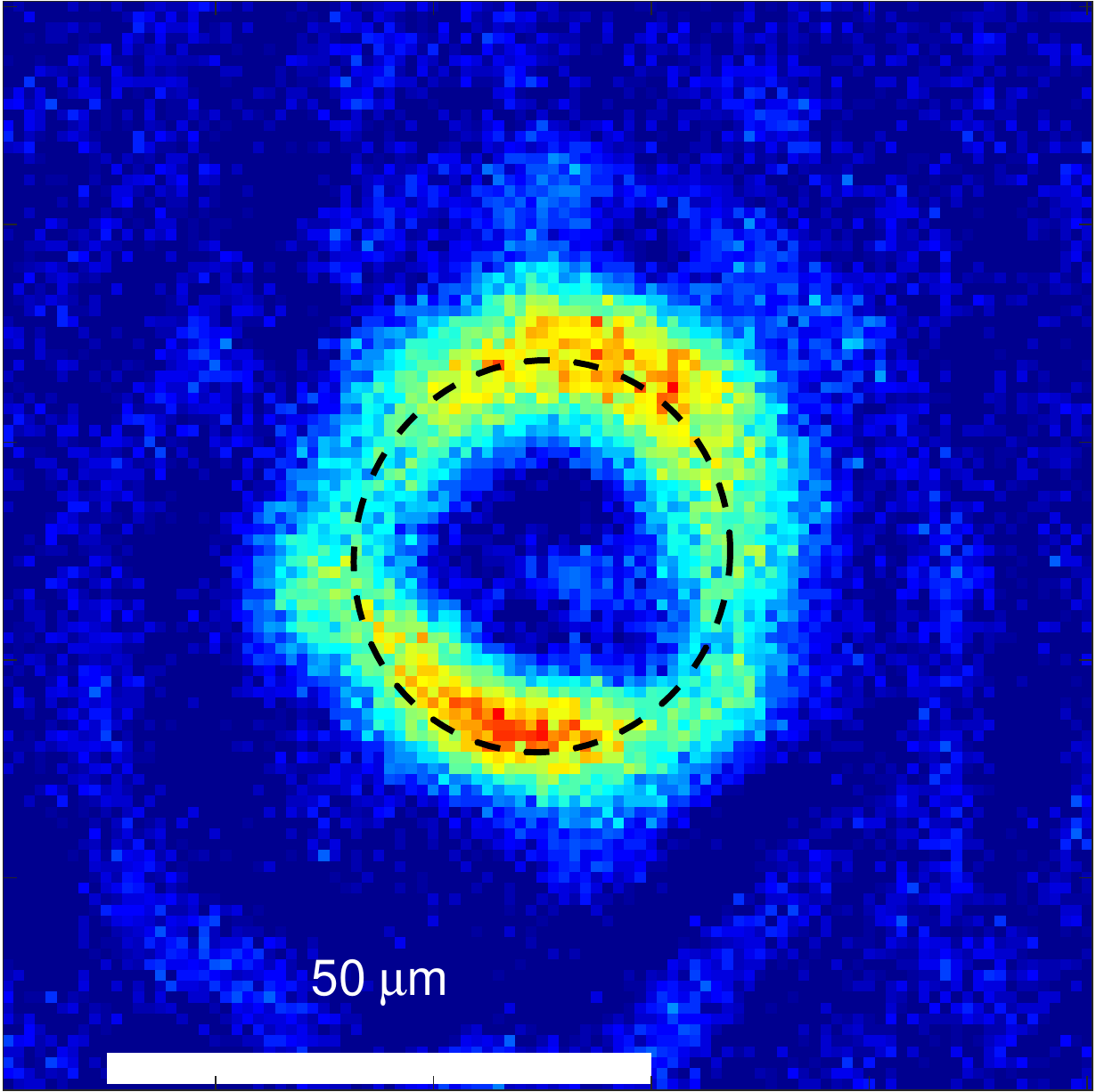}\\
(a)
\end{minipage}
\begin{minipage}{0.19\linewidth}
\centering
\includegraphics[width=0.95\linewidth]{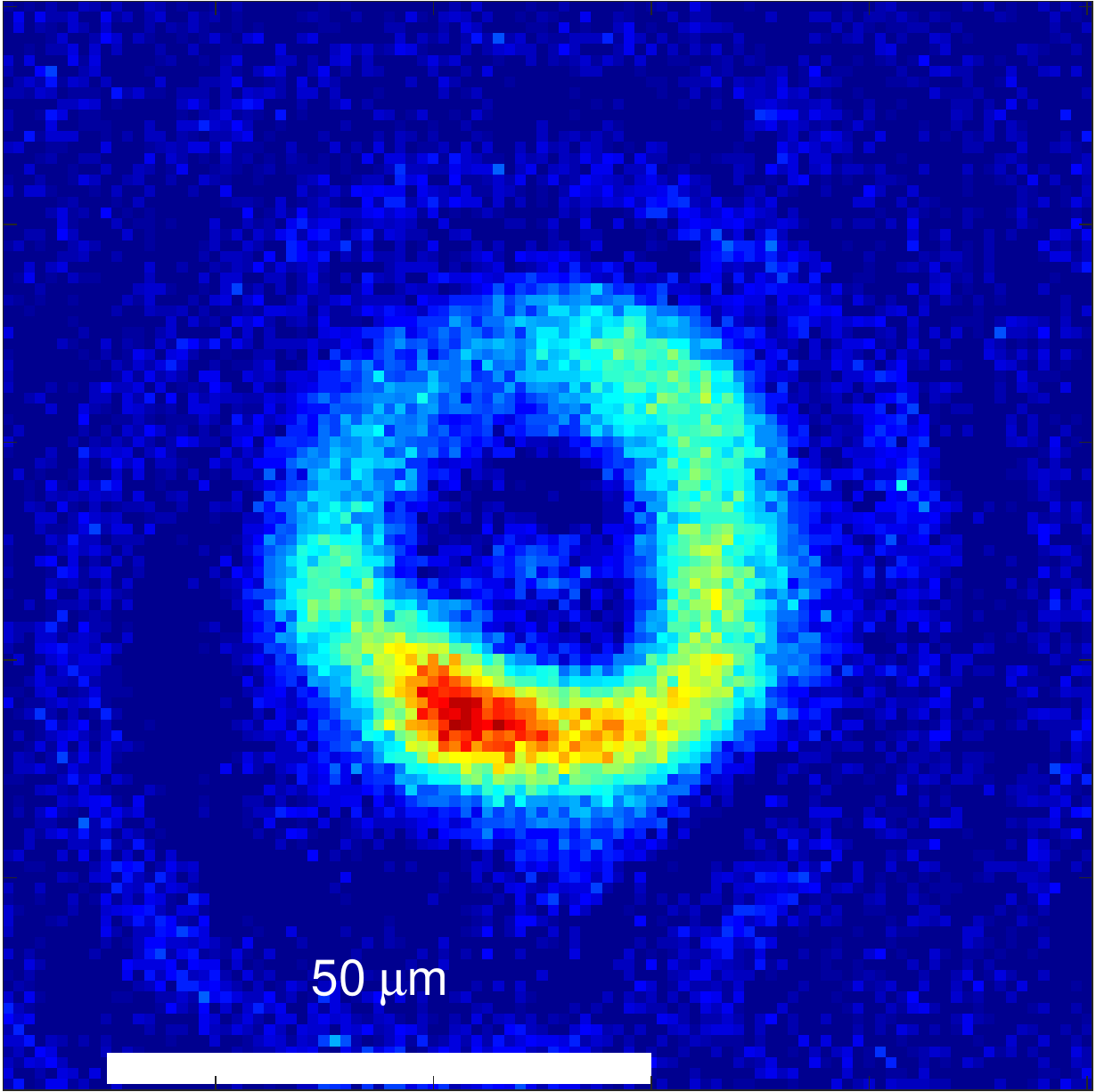}\\
(b)
\end{minipage}
\begin{minipage}{0.19\linewidth}
\centering
\includegraphics[width=0.95\linewidth]{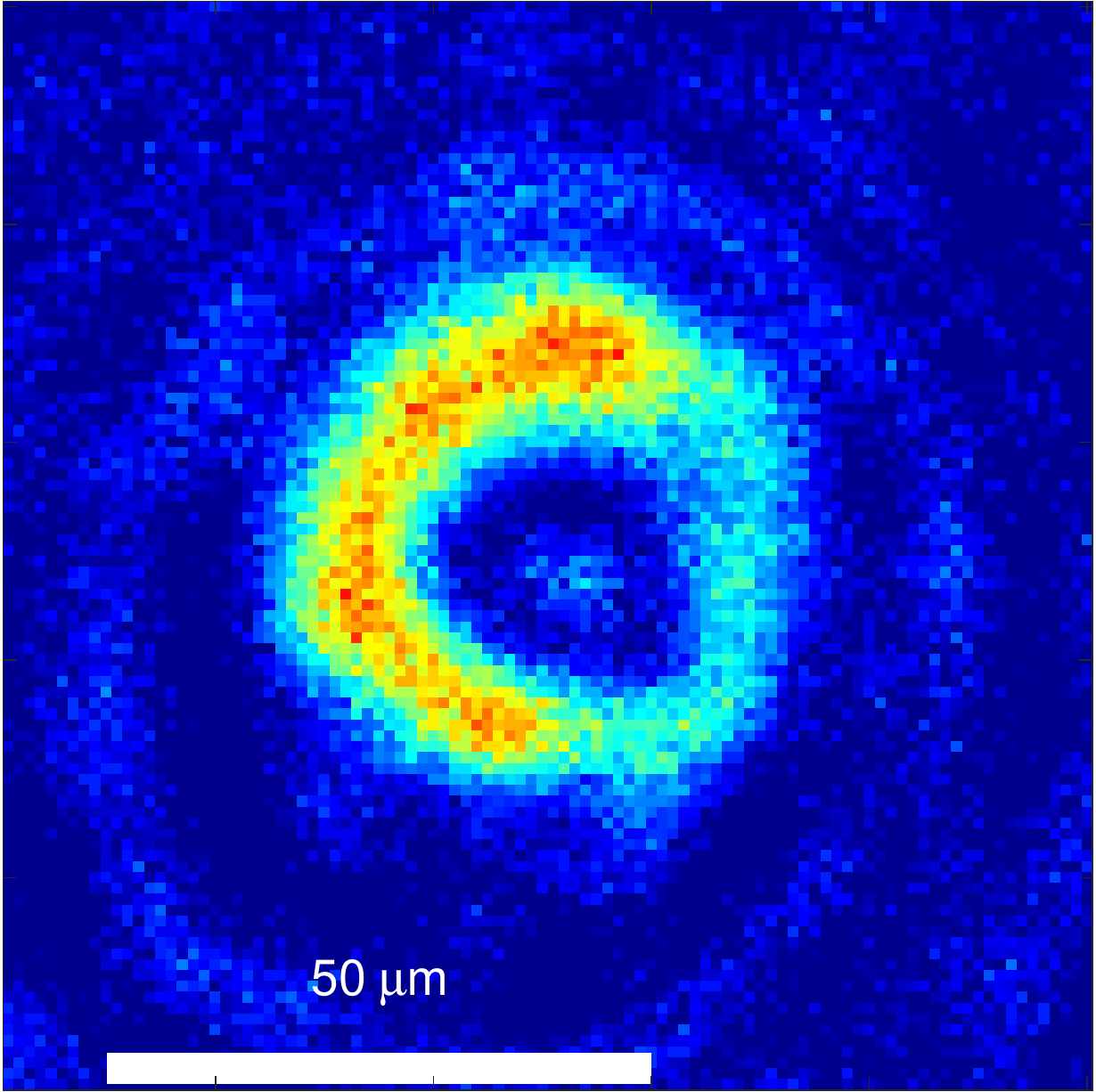}\\
(c)
\end{minipage}
\begin{minipage}{0.19\linewidth}
\centering
\includegraphics[width=0.95\linewidth]{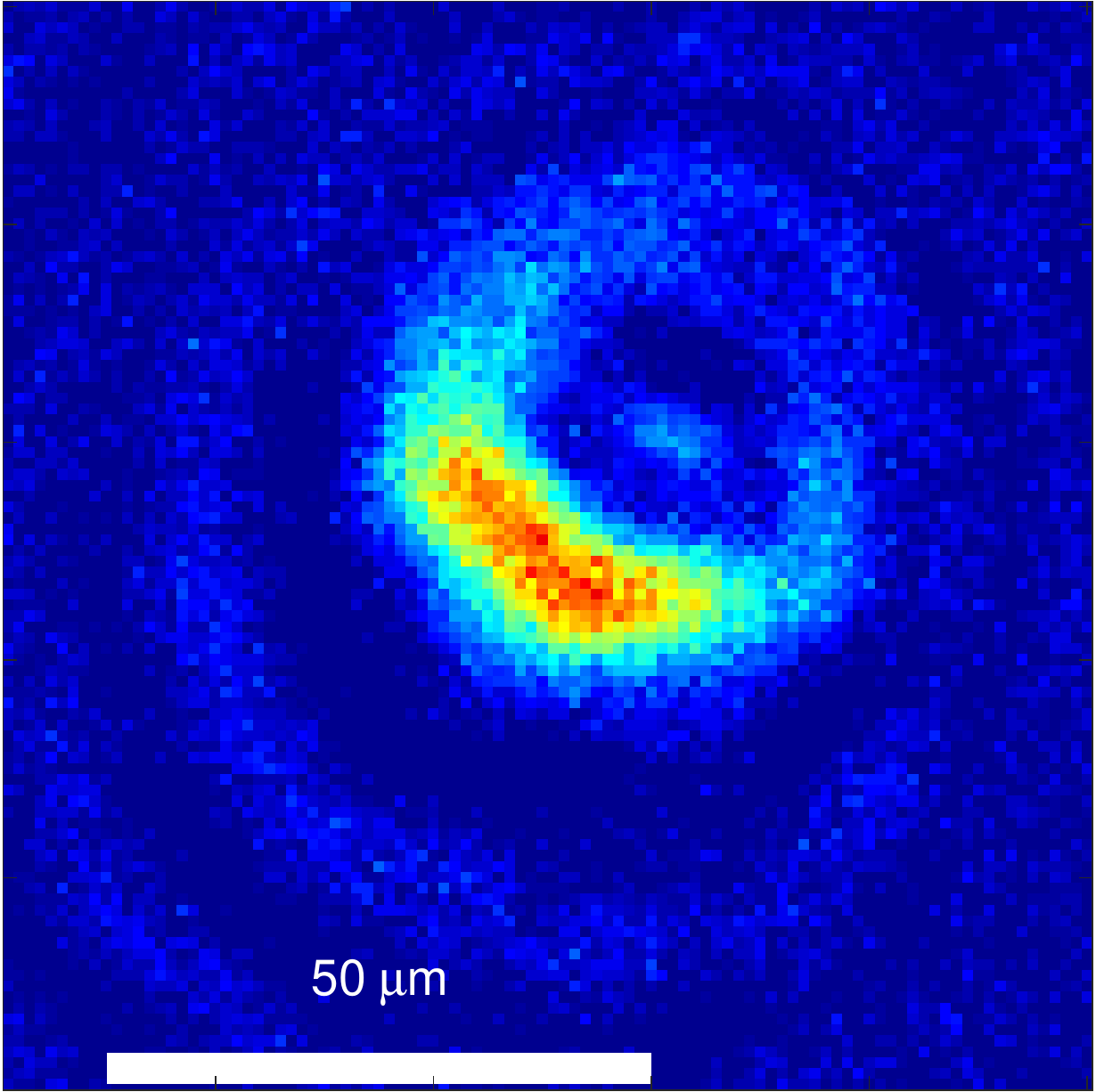}\\
(d)
\end{minipage}
\begin{minipage}{0.19\linewidth}
\centering
\includegraphics[width=0.95\linewidth]{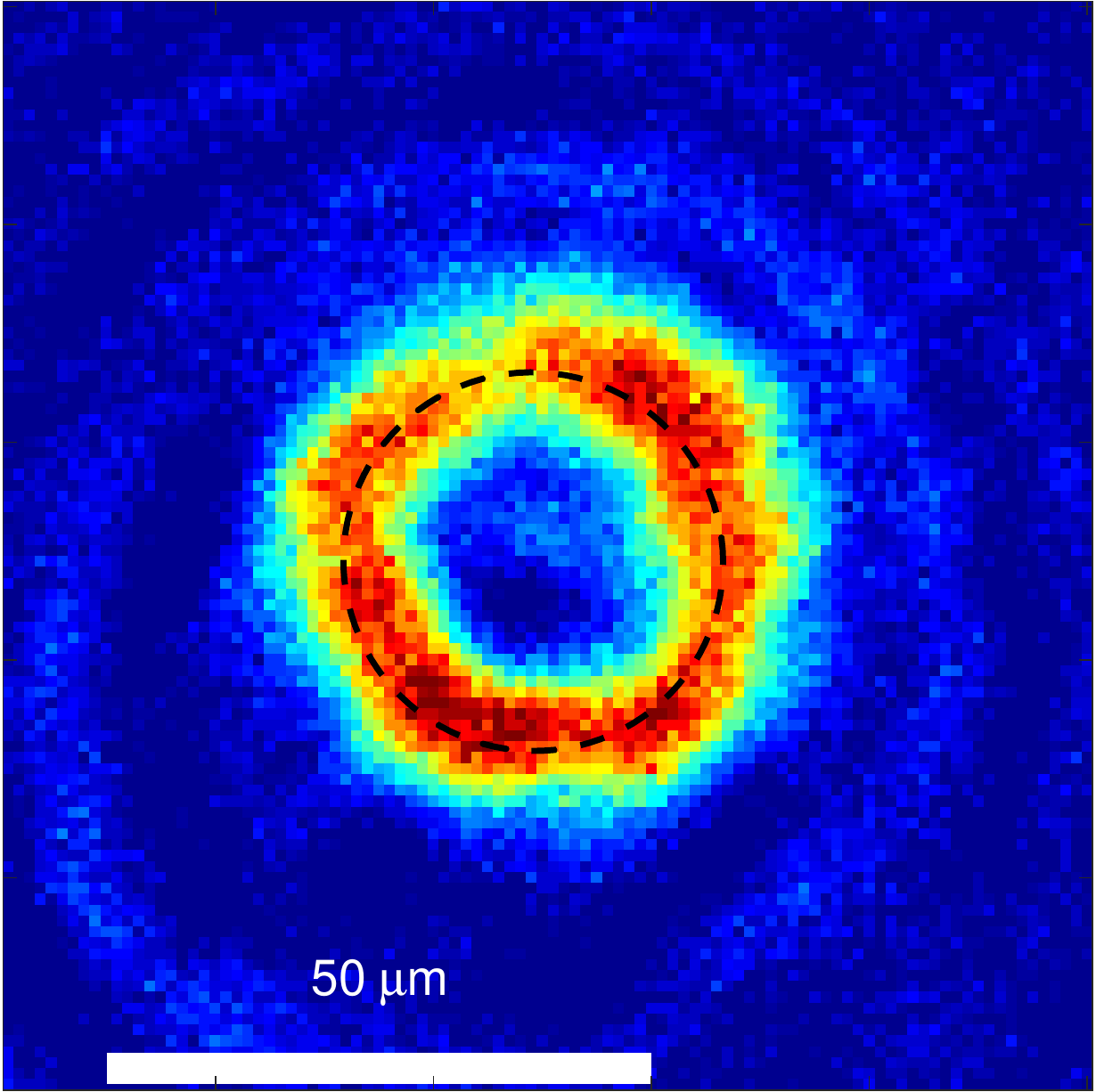}\\
(e)
\end{minipage}
\caption{Effect of the third antenna on the annular density. (a-d): Annular gas obtained at fixed rf amplitude for a relative phase of the third antenna with respect to the first antenna of (a) 0, (b) $-\pi/4$, (c) $\pi/4$, (d) $\pi$. (a): double-moon asymmetry; (d): lateral asymmetry; (b-c): combination of both; (e): fully optimized ring. These rings are achieved with a dressing frequency of \SI{300}{\kilo\hertz} and \SI{110}{\ampere} in the quadrupole coils, corresponding to a measured radius of \SI{18}{\micro\meter}. The white bar in all images has a length of \SI{50}{\micro\meter}.
}
\label{fig:third_antenna}
\end{center}
\end{figure}

The rf polarization is optimized as follows. We first produce a field with the horizontal antennas and a polarization close to circular in the horizontal plane by optimizing the isotropy of an atomic cloud at the bottom of the dressed trap. We then load the atoms into the ring trap and tune finely the phase and amplitude of the three antennas to optimize the uniformity of the atomic density along the ring. In practice, we observe two kinds of imbalance (Fig.~\ref{fig:third_antenna}): (i) A `double-moon', $\pi$-periodic imbalance coming from the ellipticity of the rf field created by the two horizontal antennas, which can be corrected by adjusting their relative phase and amplitude; (ii) and a lateral, $2\pi$-periodic imbalance appearing when the rf field created by the two main antennas is not strictly orthogonal to the vertical axis of the quadrupole coils. It is corrected by adjusting the amplitude and phase of the rf field created by the third antenna. The phase controls the azimuthal position of the density maximum.

\section{Computation of tree-body losses in the ring trap in 3D and 1D}
\label{appendix:3D_losses}

\subsection{Computation in the 3D regime}

For computing the three-body losses from a Bose-Einstein condensate confined in the ring trap in the 3D regime, we suppose that the cloud has a Thomas-Fermi profile along the $r$ and $z$ directions, and a radially uniform density profile. The atomic density then reads:
\begin{equation}
    n(\GG{r}) = n(0)\left(1-\ds\frac{(r-r_0)^2}{R_x^2}-\ds\frac{z^2}{R_z^2}\right), 
\end{equation}
where $R_i = \sqrt{2\mu\ind{3D}/M\omega_i^2}$ is the Thomas-Fermi radius along $i=r$ or $z$, and the central density is then $n(0)=\mu\ind{3D}/g=N/(\pi^2 R_r R_z r_0)$, where the 3D chemical potential is equal to \cite{Morizot2006}
\begin{equation}
\mu\ind{3D}=\hbar\overline{\omega}\sqrt{\ds\frac{2 N a}{\pi r_{0}}}.
\end{equation}

The cubic density integrated on the whole cloud volume is therefore equal to:
\begin{eqnarray}
   && \int n(\GG{r})^3 dV =\ldots\nonumber\\
    &=& n(0)^3 \times 2\pi \int_{-R_r}^{R_r} (r_0+r)\,dr \int_{-R_z\sqrt{1-\frac{r^2}{R_r^2}}}^{R_z\sqrt{1-\frac{r^2}{R_r^2}}}\left[1-\ds\frac{r^2}{R_r^2}-\frac{z^2}{R_z^2}\right]^3\nonumber\\
    & =& n(0)^3 \times 2\pi r_0 R_r R_z \int_{-1}^{1}du\int_{-\sqrt{1-u^2}}^{\sqrt{1-u^2}}dv\,(1-u^2-v^2)^3 \nonumber\\
    & =& 2\pi n(0)^3 r_0 R_r R_z \int_{0}^{1}\int_{0}^{2\pi}(1-\rho^2)^3\rho d\rho d\theta\nonumber\\
    & =& \ds\frac{\pi^2 R_r R_z}{2} n(0)^3 = \frac{N n(0)^2}{2}, 
\end{eqnarray}
and the evolution of the atom number in the cloud due to three-body recombination therefore follows

\begin{eqnarray}
    \left.\Derivee{N}{t}\right|_{3b}&=&-K_{3}\int dV n(t)^{3}\nonumber\\
    &=&-K_{3}\left(\ds\frac{M\omega_{r}\omega_{z}}{4\pi^{2} r_{0}g}\right)N^2(t).
\end{eqnarray}
Solving this differential equation while neglecting other atom loss sources therefore gives the time evolution of the atom number:
\begin{equation}
N(t)=\ds\frac{1}{\ds K_{3}\left(\frac{M\omega_{r}\omega_{z}}{4\pi^{2} r_{0}g}\right)t+\frac{1}{N(0)}}.
\end{equation}

\subsection{Computation in the 1D regime}

For computing the three-body losses in the 1D regime, we suppose this time that all atoms are in the transverse ground state. We model the ring as a 1D box with total length $2\pi r_0$, with a transverse wavefunction equal to:
\begin{equation}
    \phi_0(r,z) = \left(\ds\frac{M\ombar}{\pi\hbar}\right)^{1/2}e^{-\frac{M\omega_r}{2\hbar}r^2}e^{-\frac{M\omega_z}{2\hbar}z^2}.
\end{equation}
The atomic density profile then reads:
\begin{equation}
    n(\GG{r})=\ds\frac{N}{2\pi r_{0}}\left|\phi_{0}(r,z)\right|^2=\ds\frac{N}{2\pi r_{0}}\left(\ds\frac{M\overline\omega}{\pi\hbar}\right)e^{-\frac{M\omega_r}{\hbar}r^2}e^{-\frac{M\omega_z}{\hbar}z^2},
\end{equation}
leading to a time evolution of atom number described by:
\begin{eqnarray}
    \left.\Derivee{N(t)}{t}\right|_{3b}&=&-K_{3}\int dV n(t)^{3}\nonumber\\
    &=&-K_{3}\left(\ds\frac{N(t)}{2\pi r_{0}}\right)^3\left(\ds\frac{M\overline\omega}{\pi\hbar}\right)^3\times 2\pi r_0\ds\frac{\hbar}{3M\ombar}\frac{1}{\pi} \nonumber\\
    &=&-\ds\frac{K_3}{3}\left(\ds\frac{M\ombar}{2\pi^3 \hbar r_0}\right)^2\times N(t)^3.
\end{eqnarray}

One may note that the atom losses are, this time, proportional to the cube of number of atoms, because the repulsive interactions are not able to enlarge the atomic density profile anymore compared to the Thomas-Fermi regime.

From this differential equation, one can then compute the evolution of the atom number, again neglecting anything but three-body losses:
\begin{equation}
N(t)=\frac{1}{\sqrt{\displaystyle \frac{2}{3}K_{3} \left(\frac{M\overline{\omega}}{2\pi^3 \hbar r_{0}}\right)^2 t+\frac{1}{N^2(0)}}}.
\end{equation}

\vspace{5mm}
~

\end{document}